\documentclass[a4paper,fleqn,usenatbib]{mnras}

\usepackage[T1]{fontenc}
\usepackage{ae,aecompl}
\usepackage{newtxtext,newtxmath}

\usepackage{graphicx}	
\usepackage{xcolor}     
\usepackage{soul}

\title[Disk outflows from BH-NS mergers]{The landscape of disk outflows from black hole - neutron star mergers}

\author[Fern\'andez, Foucart, \& Lippuner]{
Rodrigo Fern\'andez$^{1}$\thanks{E-mail: rafernan@ualberta.ca}, Francois Foucart$^{2}$, Jonas Lippuner$^{3,4}$
\\
$^{1}$Department of Physics, University of Alberta, Edmonton, AB T6G 2E1, Canada\\
$^{2}$Department of Physics and Astronomy, University of New Hampshire, Durham, NH 03824, USA\\
$^{3}$CCS-2, Los Alamos National Laboratory, Los Alamos, NM 87545, USA\\
$^{4}$Center for Theoretical Astrophysics, Los Alamos National Laboratory, Los Alamos, NM 87545, USA
}

\pubyear{2020}

\begin{document}
\label{firstpage}
\pagerange{\pageref{firstpage}--\pageref{lastpage}}
\maketitle

\begin{abstract}
We investigate mass ejection from accretion disks formed in mergers of black holes (BHs)
and neutron stars (NSs). The third observing run of the LIGO/Virgo interferometers provided BH-NS
candidate events that yielded no electromagnetic (EM) counterparts.
The broad range of disk configurations expected from BH-NS mergers motivates a thorough exploration of 
parameter space to improve EM signal predictions.
Here we conduct 27 high-resolution, axisymmetric, long-term
hydrodynamic simulations of the viscous evolution of BH accretion disks that include neutrino
emission/absorption effects and post-processing with a nuclear reaction network. In the absence 
of magnetic fields, these simulations provide a lower-limit to the fraction of the initial disk mass ejected.
We find a nearly linear inverse dependence of this fraction 
on disk compactness (BH mass over initial disk radius). The dependence is related to 
the fraction of the disk mass accreted before 
the outflow is launched, which
depends on the disk position relative to the innermost stable circular orbit. We also characterize a 
trend of decreasing ejected fraction and decreasing lanthanide/actinide content with 
increasing disk mass at fixed BH mass. This trend results from  a longer time to reach weak
freezout and an increasingly dominant role of neutrino absorption at higher disk masses.
We estimate the radioactive luminosity from the disk outflow alone
available to power kilonovae over the range of configurations studied,
finding a spread of two orders of magnitude. For most of the BH-NS parameter space,
the disk outflow contribution is well below the kilonova mass upper limits for  
GW190814.
\end{abstract}

\begin{keywords}
accretion, accretion disks -- dense matter -- gravitational waves
          -- hydrodynamics -- neutrinos -- nuclear reactions, nucleosynthesis, abundances
\end{keywords}

%%%%%%%%%%%%%%%%%%%%%%%%%%%%%%%%%%%%%%%%%%%%%%%%%%

\section{Introduction}

The Advanced LIGO interferometer has completed three observing
runs -- with Advanced Virgo joining part of the way -- resulting
in the official detection of 11 binary black hole (BH) mergers
and two neutron star (NS) mergers 
\citep{ligo_gwtc-1,ligo_gw190425,ligo_gw190412}, with many more
in candidate status at the time of this writing. 
The increased sensitivity of the third observing run
also yielded an event that can be either a BH-BH or a BH-NS merger (GW190814)
(e.g., \citealt{ligo_gw190814,coughlin_2020,ackley_2020,andreoni_2020,vieira_2020,thakur_2020}). 
Only one of these
events (GW170817), however, has had electromagnetic (EM) counterparts 
detected \citep{ligo_gw170817_multi-messenger}.
While multiple
reasons can account for the lack of an EM detection (such as
a large distance, large localization area, galactic extinction, 
or Sun constraints; e.g. \citealt{foley_2020}),
the possibility remains that these sources were intrinsically fainter
than the kilonova from GW170817.

BH-NS mergers can lead to a wide range of ejected masses depending
on whether the NS is tidally disrupted by the BH. The outcome depends
on the masses of the ingoing BH and NS, as well as on the spin of the
BH and the compactness of the NS (e.g., \citealt{foucart_2018b}). The dynamical
ejecta emerges as a very neutron-rich, equatorial tidal tail that 
quickly leaves the system. The nucleosynthesis properties and
contribution to the kilonova transient are mostly set at the time
of ejection (e.g., \citealt{roberts2017}), and its properties can 
be parameterized by direct comparison with dynamical merger simulations 
(e.g., \citealt{kawaguchi_2016,kruger_2020}).

The accretion disk, on the other hand, ejects mass on a longer timescale, as
angular momentum is transported initially by gravitational torques
and later by magnetohydrodynamic (MHD) turbulence (see, e.g., 
\citealt{FM16,shibata_2019} for reviews). 
This longer evolutionary timescale
allows weak interactions to modify the composition, resulting in
a different $r$-process yield (and possibly a different kilonova color) 
than the dynamical ejecta. The
complexity of this evolution makes it very expensive to realistically
model the disk, however. Only a handful of three-dimensional general-relativistic (GR) MHD
simulations of disks around BHs 
have been carried out including at least some important microphysics or neutrino effects
\citep{siegel_2017a,nouri_2018,siegel_2018,F19_grmhd,miller2019,christie2019}.
Furthermore, all of these simulations either focus on a narrow subset of parameter
space and/or do not evolve the system for long enough to achieve completion of mass ejection.

More extensive studies of BH accretion disks have been carried out 
using axisymmetric hydrodynamic simulations with a wide variety
of approximations to the physics \citep{FM13,just2015,FKMQ14,fujibayashi2020}. 
None of these studies covers a significant fraction of all the possible
BH accretion disk configurations, however. 

Despite missing the magnetic field, hydrodynamic simulations can provide a good
description of the late-time thermal component of the outflow that arises when 
weak interactions freeze out \citep{Metzger+09a} and heating of the disk 
by viscous stresses (in lieu of MHD turbulence) is unbalanced. Close comparison between 
GRMHD and hydrodynamic simulations show that the latter provide a lower limit to the 
fraction of the disk ejected, with magnetic enhancements dependent on the strength 
of the initial poloidal field in the disk \citep{F19_grmhd,christie2019}.

Here we carry out an extensive set of long-term hydrodynamic simulations
of accretion disks around BH remnants, with the aim of sampling the entirety of the
parameter space resulting from BH-NS mergers, and thereby improving
parameter estimation models that take disk outflow properties as input 
(e.g., \citealt{barbieri_2019,hinderer_2019,coughlin_2020}).
Along the way, a broad probe of parameter space allows to identify trends 
in the disk ejection physics, helping to focus on areas where improvements in
the physics (e.g. neutrino transport) have the most impact.
Finally, by providing a lower limit to the disk mass ejected, we are
also estimating the lower limit to the raw radioactive heating available to
power kilonova transients.

The structure of the paper is as follows. Section \ref{s:methods} presents our methods,
including our choice of initial conditions from the plausible
parameter space of BH-NS mergers. Section \ref{s:results} presents our results, divided
into mass ejection, outflow composition, and implications for EM
counterparts. We close with a summary and discussion in Section \ref{s:summary}.

%%%%%%%%%%%%%%%%%%%%%%%%%%%%%%%%%%%%%%%%%%%%%%%%%%%%%%%%%%%%%%%%%%%%%%%%%%
\section{Methods}
\label{s:methods}

\subsection{Initial conditions}
\label{s:initial_conditions}

\begin{figure*}
\includegraphics*[width=\textwidth]{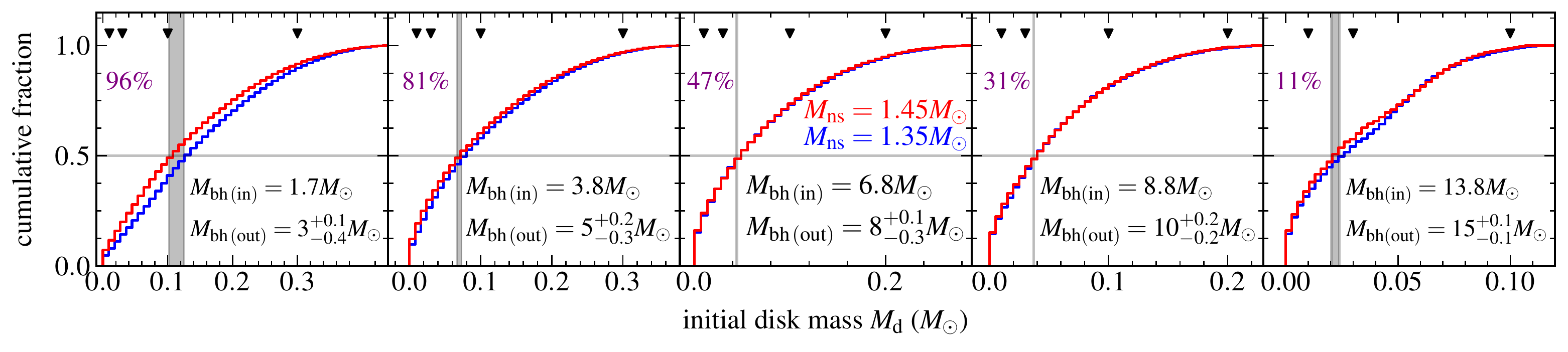}
\caption{Cumulative distribution of post-merger disk masses, obtained by using
analytic formulae for the post-merger BH mass and spin \citep{pannarale_2014}, remnant mass 
outside the BH \citep{foucart_2018b}, and mass in dynamical ejecta \citep{kruger_2020}. 
For fixed ingoing NS mass $M_{\rm ns}$ and BH mass $M_{\rm bh(in)}$, the intervals 
$9-13$\,km for NS radii and $0-0.7$ for ingoing BH spins are uniformly sampled. 
The uncertainty range in the post-merger BH mass $M_{\rm bh(out)}$ 
indicates the range in median
values obtained by using the NS masses shown, with the central value corresponding to 
$M_{\rm ns}=1.4M_\odot$. The triangles indicate the initial disk masses selected for
our hydrodynamic simulations (c.f. Table~\ref{t:models}). The fraction of mergers that result 
in NS disruption for each $\{M_{\rm bh(in)},M_{\rm ns}\}$ pair is shown in purple.}
\label{f:mdisk_initial}
\end{figure*}

In order to sample representative initial disk masses for our simulations, we 
map the parameter space of BH-NS merger remnants using analytical formulae that
are calibrated to numerical relativity simulations. For a given
ingoing BH mass $M_{\rm bh(in)}$ and NS mass $M_{\rm ns}$,
we uniformly sample the range $9-13$\,km for NS radii and
$0-0.7$ for the ingoing BH spin, and compute distributions of 
(1) the remnant baryon mass left outside the BH using the formula of
\citet{foucart_2018b}, (2) the disk mass using the formula of \citet{kruger_2020}
for the dynamical ejecta, and (3) the post-merger BH mass $M_{\rm bh(out)}$ and its 
spin using the formulae of \citet{pannarale_2014} and the output
from steps (1)-(2). This approach is intended to be agnostic about the 
properties of the EOS of dense matter and initial BH spin distribution,
within plausible limits.

The resulting cumulative distributions of initial disk masses are shown in Figure~\ref{f:mdisk_initial}
for two NS mases $\{1.35,1.45\}\,M_\odot$ and ingoing BH masses such that the median value
of $M_{\rm bh(out)}$ is $\{3,5,8,10,15\}\,M_\odot$. The lowest post-merger BH mass is chosen to explore 
the hypothetical case of a very low-mass ingoing BH, or the prompt collapse of a NS-NS system. 
The highest BH mass is chosen such that at least $10\%$ of mergers result in disruption. 
Median disk masses range from $0.1 M_\odot$ for the $3 M_\odot$ post-merger BH, to $0.02 M_\odot$ 
for the $15M_\odot$ BH. In most cases, sensitivity to the specific choice of NS mass does alter 
the shape of the histogram but not the extreme values. The post-merger BH spin distributions
have medians in the range $0.85-0.9$, except for the lowest mass BHs considered, which
have spins $>0.95$.

For each post-merger BH mass, Figure~\ref{f:mdisk_initial} shows the disk masses sampled in our 
study. The lowest disk mass in all cases is taken to be $0.01M_\odot$, which is optically thin 
to neutrinos, while the largest disk mass is such that it is at the 
uppermost end of plausible values. In all of our models, we take the spin of the post-merger BH to be
$0.8$, which while somewhat lower than the median values of the distributions obtained,
lies within the range of plausible values (except for $M_{\rm bh(out)}=3M_\odot$) 
and does not demand a prohibitively small time step. Furthermore, the effect of BH 
spin on the disk outflow properties is known, with more mass with higher average
electron fraction being ejected for higher BH spins \citep{FKMQ14,fujibayashi2020}.

Our simulations start from idealized equilibrium tori (\S\ref{s:flash}) which 
approach a Keplerian angular velocity distribution after a few orbits. This equilibrium configuration
requires more parameters in addition to the disk mass: a radius of maximum density, an entropy 
(internal energy content), and an electron fraction (composition). While these parameter
choices would not be necessary if we mapped the disk directly from a merger simulation,
our broad coverage of possible BH-NS combinations would be limited by accessible merger simulation data. 
The remaining initial disk parameters are therefore chosen by inspecting the output of 
numerical relativity simulations of BH-NS mergers 
and making educated guesses about these values in 
regimes not covered by simulations.

We adopt initial disk radii that roughly follow the location of 
density maxima outside of the BH in published BH-NS simulations
\citep{etienne_2009,kyutoku_2010,foucart_2011,foucart_2013,foucart_2014,
kawaguchi_2015,kyutoku_2015,foucart_2017,brege_2018}. 
This radius of maximum density is not well defined,
however, since it depends on (1) the time at which it is measured, (2) the metric, and (3) 
on whether the local density, surface density, or enclosed mass are reported. Since
there is no consistency across the literature for this quantity, 
we adopt a fiducial set of initial disk radii $R_{\rm d}=\{50,50,60,90,120\}$\,km
for post-merger BH masses $\{3,5,8,10,15\}\,M_\odot$, respectively. 

The entropy of all disks is taken to be $8\,k_{\rm B}$ per baryon, which results in ratios of isothermal sound 
speed to orbital speed $\sim 10-30\%$ at the point of maximum density. 
While this choice has some effect on the amount of mass ejected, we use a constant
value for uniformity.
Finally, the default electron fraction of the 
disks is set to $Y_{e,\rm ini} = 0.2$, although we vary this parameter in our
simulations given that it is dependent on the quality of the neutrino transport
implementation in the merger simulation, and has a non-negligible effect on the
disk outflow composition.

%----------------------------------------------------------------------------------
\subsection{Hydrodynamic Simulations}
\label{s:flash}

We perform time-dependent hydrodynamic simulations with {\tt FLASH} version 3.2
\citep{fryxell00,dubey2009}, with the modifications described in \citet{FM13},
\citet{MF14}, \citet{FKMQ14}, and \citet{lippuner_2017}.  The code solves the
Euler equations in axisymmetric spherical polar coordinates $(r,\theta)$,
subject to source terms that include the pseudo-Newtonian gravitational
potential of a spinning BH \citep{artemova1996} without disk self-gravity, shear viscosity with an
$\alpha$ parameterization \citep{shakura1973}, and a leakage scheme for
neutrino emission, with absorption included as a disk-like light bulb
\citep{FM13,MF14}. We only include electron type neutrinos/antineutrinos
interacting with nucleons via charged-current weak interactions.  The code
employs the equation of state (EOS) of \citet{timmes2000} with the abundances
of neutrons, protons, and alpha particles in nuclear statistical equilibrium
(NSE) above a temperature $T=5\times 10^9$\,K and accounting for the nuclear binding energy
of these particles. The electron-positron quantities are extended above the
high-density limit of the table using analytic expressions \citep{bludman1978,bethe80}.

The initial condition is an equilibrium torus with constant angular momentum,
entropy, and electron fraction, with mass fractions assumed to be in NSE (e.g., \citealt{FM13}).
Parameters are chosen according to \S\ref{s:initial_conditions}. The floor of density
is set to $10$\,g\,cm$^{-3}$ at $r=4R_{\rm d}$, and has an initial radial dependence $r^{-2}$. 
For $r\leq 4R_{\rm d}$, the radial exponent of the floor is smoothly brought to zero on
a timescale of $40$ orbital times at $r=R_{\rm d}$, reaching a flat floor 
in this region (\citealt{F19_grmhd}, see also \citealt{just2015}). 
The initial ambient density is set at $1.1$ times the floor.

The computational domain extends from an inner radius $r_{\rm in}$ midway between the radius
of the innermost stable circular orbit (ISCO) $r_{\rm isco}$
and the BH horizon, to an outer radius $r_{\rm out} = 10^4 r_{\rm in}$, with the
polar angle spanning the range $[0,\pi]$. The grid is discretized logarithmically
in radius, using $128$ cells per decade, and a polar grid equispaced in
$\cos\theta$ using $112$ cells. On the equatorial plane, this results in 
a spacing $\Delta r /r \simeq 1.8\% \simeq 1^\circ \simeq \Delta \theta$. This
resolution is double that of the models in \citet{fernandez_2017},
equivalent to that of the high-resolution models of \citet{FM13} and 
\citet{FKMQ14}, and the same as in \citet{fahlman_2018}
and the hydrodynamic models of \citet{F19_grmhd}. The boundary conditions are set to outflow in radius and
reflecting in $\theta$.

%----------------------------------------------------------------------------------
\subsection{Nuclear Reaction Network Post-Processing}
\label{s:skynet}

Passive tracer particles are initially placed in the disk following the density
distribution. For each hydrodynamic simulation we employ $10^4$ particles, each
representing an equal amount of mass. Particles are advected with the flow and record 
various kinematic and thermodynamic quantities 
as a function of time. Particles that are ejected with positive Bernoulli parameter beyond a radius of
$10^9$\,cm by the end of the simulation are considered to be part of the disk
outflow.

Outflow trajectories are post-processed with the nuclear reaction network {\tt SkyNet} \citep{lippuner_skynet}, 
using the same settings as in \citet{lippuner_2017}.  The network employs 7843 nuclides
and more than $1.4\times 10^5$ reactions, including strong forward reaction rates from the
REACLIB database \citep{cyburt_2010} with inverse rates computed from detailed balance;
spontaneous and neutron-induced fission rates from \citet{frankel_1947}, \citet{mamdouh_2001},
\citet{wahl_2002}, and \citet{panov_2010}; weak rates from \citet{fuller_1982}, \citet{oda_1994},
\citet{langanke_2000}, and the REACLIB database; and nuclear masses from the REACLIB
database, which includes experimental values were available, or otherwise theoretical masses
from the finite-range droplet macroscopic model (FRDM) of \citet{moeller_2016}.

The rates of electron neutrino/antineutrino absorption/emission recorded by the trajectory 
are included in the evolution of the proton and neutron fraction. Likewise, the
temperature and entropy are evolved self-consistently by accounting for nuclear heating
from the network, as well as viscous heating and neutrino heating/cooling in the hydrodynamic
simulation as recorded by the trajectory.

Processing begins when the trajectory reaches $10$\,GK for the last time,
or when the temperature is maximal if lower than $10$\,GK at all times. For the portion of the
evolution in which the temperature is higher than $7$\,GK, abundances are evolved in 
NSE, subject to neutrino interactions, while full network integration is carried out at lower temperatures.
Trajectories are extended beyond the end of the simulation
($12-25$\,s) by assuming that the density decays with time as
$t^{-3}$, to allow $r$-process nuclei with long half-lives to decay. 
Since the $r$-process is complete by the time this transition is made, most of the nuclear heating
has already been deposited and the exact time dependence of the density decay is not important. 
While trajectories are evolved until $30$\,yr, information is
extracted at $t=1$\,day and $t=1$ week to estimate the properties of the kilonova
at peak.

%-----------------------------------------------------------------------------------
\subsection{Models Evolved}
\label{s:models}

Table~\ref{t:models} shows all of the hydrodynamic models we evolve. As a baseline set,
we take disks with initial conditions as described in \S\ref{s:initial_conditions}:
black hole and disk masses as in Figure~\ref{f:mdisk_initial}, initial entropy
$8k_{\rm B}$ per baryon, and initial electron fraction $Y_{\rm e, ini}= 0.2$.
Model names follow the convention bXXdYY, where XX and YY refer to the BH mass
and disk mass, respectively.
Tori are constructed as an equilibrium solution to the momentum equation with
constant Bernoulli parameter, constant angular momentum, and the pseudo-Newtonian
potential of the BH \citep{FM13}. The torus shape is controlled by a dimensionless
distortion parameter $d$ \citep{stone1999} which is solved for by fixing the entropy, $Y_e$, and torus mass.
The distortion parameter is related to the torus initial Bernoulli parameter $b_{\rm ini}$,
black hole mass $M_{\rm bh}\equiv M_{\rm bh(out)}$, and radius of initial 
density peak $R_{\rm d}$ by 
\begin{equation}
\label{eq:berc_disk}
b_{\rm ini} = -\frac{1}{2d}\frac{GM_{\rm bh}}{R_d}.
\end{equation}
The baseline set is evolved with a viscosity parameter $\alpha=0.03$

To assess the effects of initial composition, we evolve a few models with 
lower initial electron fraction than the baseline set $Y_{\rm e, ini}= \{0.10,0.15\}$.
Likewise, we evolve two models with higher viscosity parameter, $\alpha=0.1$. 

All hydrodynamic models are evolved for $5,000$ orbits at the initial density peak radius,
which corresponds to $\simeq 12-25$\,s of physical time (Table~\ref{t:models}). 
This time is chosen such that the mass ejection from the disk is mostly complete.
Tracer particles from each simulation are then post-processed with the nuclear reaction
network as described in \S\ref{s:skynet}

\begin{table}
\centering
\caption{Hydrodynamic models evolved and input parameters. Columns from left to right show model name,
black hole mass, disk mass, radius of initial disk density peak, initial electron fraction, torus
distortion parameter, viscosity parameter, and maximum evolution time.}
\setlength\tabcolsep{5.5pt}
\label{t:models}
\begin{tabular}{lccccccc} 
\hline
Model & $M_{\rm bh}$ & $M_{\rm d}$          & $R_{\rm d}$ & $Y_{e,\rm ini}$ & $d$ & $\alpha$ & $t_{\rm max}$\\
      & ($M_\odot$)  & ($M_\odot$) & (km)        & {}              & {} & {} & (s) \\
\hline
b03d01  & 3 & 0.01  & 50 & 0.20 & 1.52 & 0.03 & 16.5 \\
b03d03  &   & 0.03  &    &      & 1.82 &      &      \\
b03d10  &   & 0.10  &    &      & 2.40 &      &      \\
b03d30  &   & 0.30  &    &      & 3.37 &      &      \\
\noalign{\smallskip}
b05d01  & 5 & 0.01  &    &      & 1.39 &     & 12.2 \\
b05d03  &   & 0.03  &    &      & 1.59 &     &      \\
b05d10  &   & 0.10  &    &      & 1.97 &     &      \\
b05d30  &   & 0.30  &    &      & 2.57 &     &      \\
\noalign{\smallskip}
b08d01  & 8 & 0.01  & 60 &      & 1.30 &    & 12.1 \\
b08d03  &   & 0.03  &    &      & 1.43 &    &      \\
b08d10  &   & 0.10  &    &      & 1.67 &    &      \\
b08d20  &   & 0.20  &    &      & 1.87 &    &      \\
\noalign{\smallskip}
b10d01  & 10 & 0.01  & 90 &     & 1.20 &    & 20.6 \\
b10d03  &    & 0.03  &    &     & 1.30 &    &      \\
b10d10  &    & 0.10  &    &     & 1.46 &    &      \\
b10d20  &    & 0.20  &    &     & 1.59 &    &      \\
\noalign{\smallskip}
b15d01  & 15 & 0.01  & 120 &    & 1.15 &    & 25.4 \\
b15d03  &    & 0.03  &     &    & 1.22 &    &      \\
b15d10  &    & 0.10  &     &    & 1.32 &    &      \\
\noalign{\smallskip}
\noalign{\smallskip}
b03d01-y10 & 3 & 0.01  & 50 & 0.10 & 1.50 &   & 16.5 \\
b03d30-y10 &   & 0.30  &    &      & 3.30 &   &      \\
b08d03-y10 & 8 & 0.03  & 60 &      & 1.42 &   & 12.1 \\
\noalign{\smallskip}                      
b03d01-y15 & 3 & 0.01  & 50 & 0.15 & 1.51 &   & 16.5 \\
b03d30-y15 &   & 0.30  &    &      & 3.32 &   &      \\
b08d03-y15 & 8 & 0.03  & 60 &      & 1.43 &   & 12.1 \\
\noalign{\smallskip}
\noalign{\smallskip}
b03d01-v10  & 3 & 0.01 & 50 & 0.20 & 1.52 & 0.10 & 16.5 \\
b08d03-v10  & 8 & 0.03 & 60 &      & 1.43 &      & 12.1 \\
\hline
\end{tabular}
\setlength\tabcolsep{6pt}
\end{table}

%%%%%%%%%%%%%%%%%%%%%%%%%%%%%%%%%%%%%%%%%%%%%%%%%%%%%%%%%%%%%%%%%%%%%%%%%%%%
\section{Results}
\label{s:results}

The overall evolution of neutrino cooled accretion disks follows well-known stages 
(e.g., \citealt{ruffert1999,popham1999,DiMatteo+02,setiawan2006,Chen&Beloborodov07,Lee+09}).
Depending on the initial disk mass, neutrinos can be trapped or escape freely. In the former case,
an initial optically thick phase ensues until the density has decreased sufficiently for transparency.
Thereafter, neutrino cooling is important compared to viscous heating,
the inner disk is not too thick vertically, and accretion proceeds efficiently. After
about a viscous time $R_{\rm d}^2/(\alpha c_i^2/\Omega_{\rm K})\sim $ few $100$\,ms (with 
$c_i$ the isothermal sound speed and $\Omega_{\rm K}$ the orbital frequency; \citealt{shakura1973}), 
the density becomes low enough that weak interactions freeze out, shutting down 
cooling (e.g., \citealt{Metzger+09a}). At this point, the disk is radiatively inefficient, 
with viscous heating and nuclear recombination of $\alpha$ particles being unbalanced, 
thus an outflow is launched until no more mass is available to be ejected.

In the absence of magnetic fields, this thermal outflow is the only relevant
mass ejection channel when a BH sits at the center, as neutrino-driven winds are weak given that self-irradiation
is not efficient (e.g., \citealt{just2015}). A comparison with long-term GRMHD simulations shows that
the outflow from hydrodynamic simulations is of similar quantity and has similar velocities
as the analog process occurring due to dissipation of MHD turbulence \citep{F19_grmhd}. 
The magnetic field provides for additional, faster components that can eject a comparable
amount of mass than the thermal outflow.

In the following, we discuss mass ejection properties across the range of models we evolve,
the composition of these ouflows, and the implications for EM counterparts of BH-NS mergers.

\begin{table*}
\begin{minipage}{16.5cm}
\centering
\caption{Summary of results. Columns from left to right show model name,
disk compactness (eq.~\ref{eq:disk_compactness_def}),
ejected mass, fraction of initial disk mass ejected $M_{\rm ej}/M_{\rm d}$,
average ouflow velocity, average outflow electron fraction, 
ejecta mass with $X_{\rm La+Ac}< 10^{-4}$ ($M_{\rm blue}$), 
ejecta mass with $X_{\rm La+Ac}> \{10^{-4},10^{-3},10^{-2}\}$ ($\{M_{-4},M_{-3},M_{-2}\}$, respectively),
and radioactive heating power (in units of $10^{40}$\,erg\,s$^{-1}$) at 1 day and 7 days, ignoring
thermalization efficiency.}
\label{t:results}
\begin{tabular}{lccccccccccc} 
\hline
Model & $C_{\rm d}$ & $M_{\rm ej}$ & $M_{\rm ej}/M_{\rm d}$ & $\langle v/c\rangle$ & $\langle Y_e\rangle$  & 
      $M_{\rm blue}/M_{\rm ej}$ & $M_{-4}/M_{\rm ej}$ & $M_{-3}/M_{\rm ej}$ & $M_{-2}/M_{\rm ej}$ 
      & $L_{\rm 40,1d}$ & $L_{\rm 40,1w}$ \\
      & & ($10^{-2}\,M_\odot$) & {} & {($10^{-2}$)} & {}  & {} & {} & {}& {} & {} & {} \\
\hline
b03d01 & 0.60 & 0.21 & 0.21 & 3.4 & 0.28 & 0.75 & 0.25 & 0.18 & 0.10 & 8.9  & 0.99 \\
b03d03 &      & 0.57 & 0.19 & 3.3 & 0.28 & 0.84 & 0.16 & 0.09 & 0.03 & 21   & 2.2  \\
b03d10 &      & 1.8  & 0.18 & 3.4 & 0.29 & 0.93 & 0.07 & 0.03 & 0.01 & 60   & 5.9\\
b03d30 &      & 4.8  & 0.16 & 3.1 & 0.29 & 0.86 & 0.14 & 0.08 & 0.05 & 150  & 17\\
\noalign{\smallskip}                        
b05d01 & 1.00 & 0.11 & 0.11 & 3.4 & 0.29 & 0.72 & 0.28 & 0.20 & 0.09 & 4.5 & 0.51\\
b05d03 &      & 0.32 & 0.11 & 3.4 & 0.30 & 0.87 & 0.13 & 0.04 & 0.02 & 12  & 1.2\\
b05d10 &      & 0.98 & 0.09 & 3.5 & 0.31 & 0.98 & 0.02 & 0.01 & 0.01 & 30  & 2.8\\
b05d30 &      & 2.7  & 0.09 & 3.1 & 0.31 & 0.98 & 0.02 & 0.02 & 0.01 & 72  & 6.3\\
\noalign{\smallskip}                        
b08d01 & 1.33 & 0.05 & 0.05 & 3.3 & 0.28 & 0.66 & 0.34 & 0.25 & 0.12 & 2.3 & 0.25\\
b08d03 &      & 0.15 & 0.05 & 3.9 & 0.30 & 0.82 & 0.18 & 0.08 & 0.05 & 5.5 & 0.58\\
b08d10 &      & 0.49 & 0.05 & 3.6 & 0.33 & 0.99 & 0.01 & 0.01 & 0.00 & 13  & 1.1\\
b08d20 &      & 0.87 & 0.04 & 3.9 & 0.33 & 1.00 & 0.00 & 0.00 & 0.00 & 19  & 1.6\\
\noalign{\smallskip}                        
b10d01 & 1.11 & 0.09 & 0.09 & 3.4 & 0.29 & 0.72 & 0.28 & 0.20 & 0.12 & 3.9 & 0.47\\
b10d03 &      & 0.27 & 0.09 & 3.4 & 0.31 & 0.86 & 0.14 & 0.06 & 0.02 & 8.6 & 0.99\\
b10d10 &      & 0.90 & 0.09 & 3.3 & 0.33 & 0.98 & 0.02 & 0.02 & 0.01 & 23  & 2.5\\
b10d20 &      & 1.7  & 0.08 & 3.5 & 0.34 & 0.99 & 0.01 & 0.01 & 0.00 & 40  & 4.3\\
\noalign{\smallskip}                        
b15d01 & 1.25 & 0.06 & 0.06 & 3.7 & 0.27 & 0.63 & 0.37 & 0.27 & 0.14 & 2.9 & 0.35\\
b15d03 &      & 0.19 & 0.06 & 3.5 & 0.30 & 0.79 & 0.21 & 0.12 & 0.04 & 6.8 & 0.80\\
b15d10 &      & 0.62 & 0.06 & 3.5 & 0.34 & 0.98 & 0.02 & 0.01 & 0.01 & 16  & 1.6\\
\noalign{\smallskip}
\noalign{\smallskip}
b03d01-y10 & 0.60 & 0.21 & 0.21 & 3.4 & 0.26 & 0.62 & 0.38 & 0.32 & 0.25 & 9.0 & 1.1\\
b03d30-y10 &      & 4.1  & 0.14 & 3.4 & 0.28 & 0.86 & 0.14 & 0.10 & 0.08 & 130 & 13\\
b08d03-y10 & 1.33 & 0.16 & 0.05 & 3.7 & 0.29 & 0.75 & 0.25 & 0.20 & 0.16 & 6.0 & 0.71\\
\noalign{\smallskip}                      
b03d01-y15 & 0.60 & 0.20 & 0.20 & 3.3 & 0.27 & 0.69 & 0.31 & 0.25 & 0.19 & 8.9 & 1.1\\
b03d30-y15 &      & 4.3  & 0.14 & 2.8 & 0.28 & 0.85 & 0.15 & 0.12 & 0.09 & 120 & 13\\
b08d03-y15 & 1.33 & 0.15 & 0.05 & 3.7 & 0.30 & 0.76 & 0.24 & 0.18 & 0.10 & 5.5 & 0.64\\
\noalign{\smallskip}                      
\noalign{\smallskip}                      
b03d03-v10 & 0.60 & 0.28 & 0.28 & 5.6 & 0.27 & 0.55 & 0.45 & 0.41 & 0.34 & 13  & 1.2 \\
b08d03-v10 & 1.33 & 0.31 & 0.10 & 5.5 & 0.31 & 0.77 & 0.23 & 0.20 & 0.15 & 12  & 0.83\\
\hline
\end{tabular}
\end{minipage}
\end{table*}

\begin{figure*}
\includegraphics*[width=0.49\textwidth]{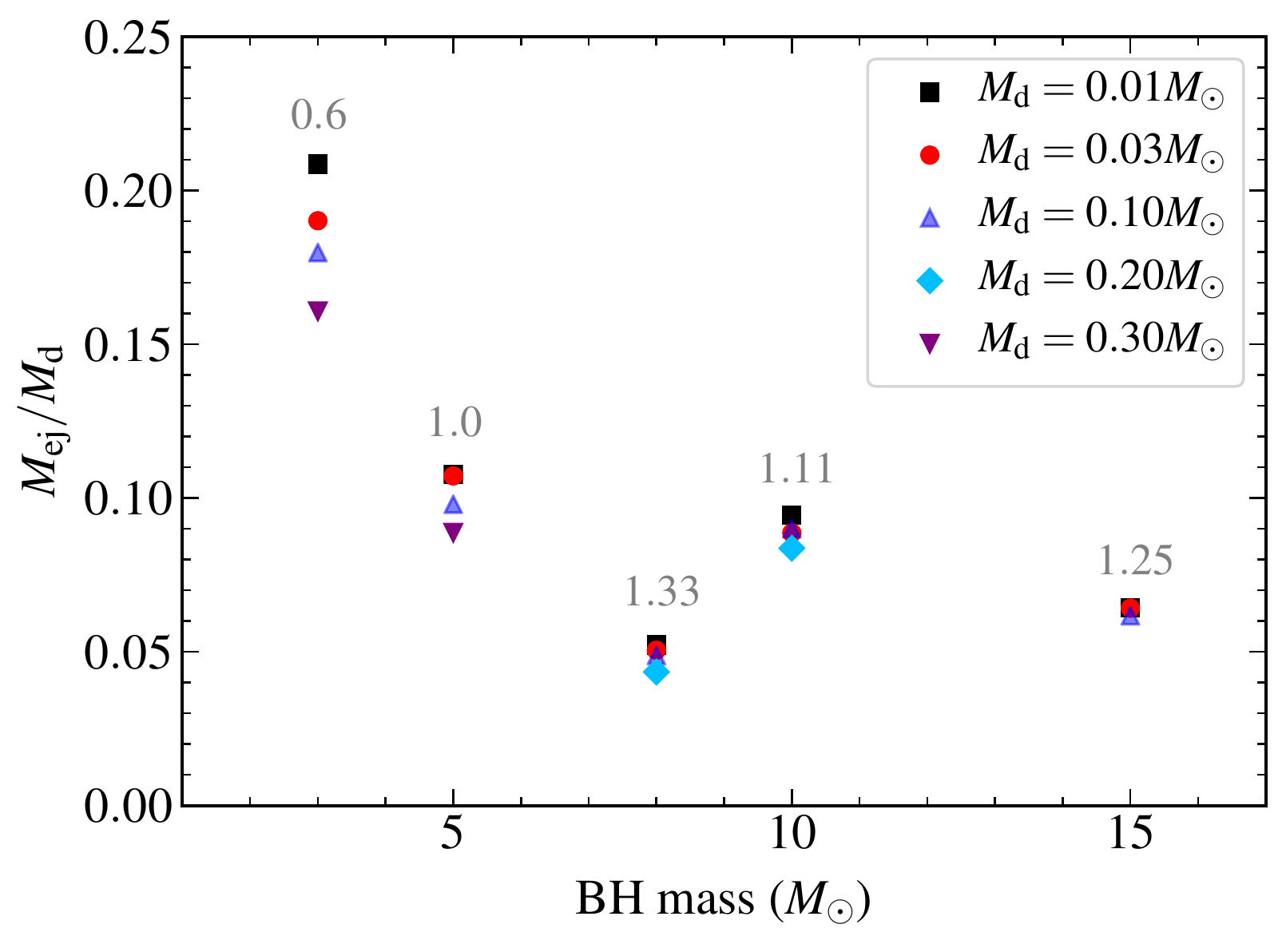}
\includegraphics*[width=0.49\textwidth]{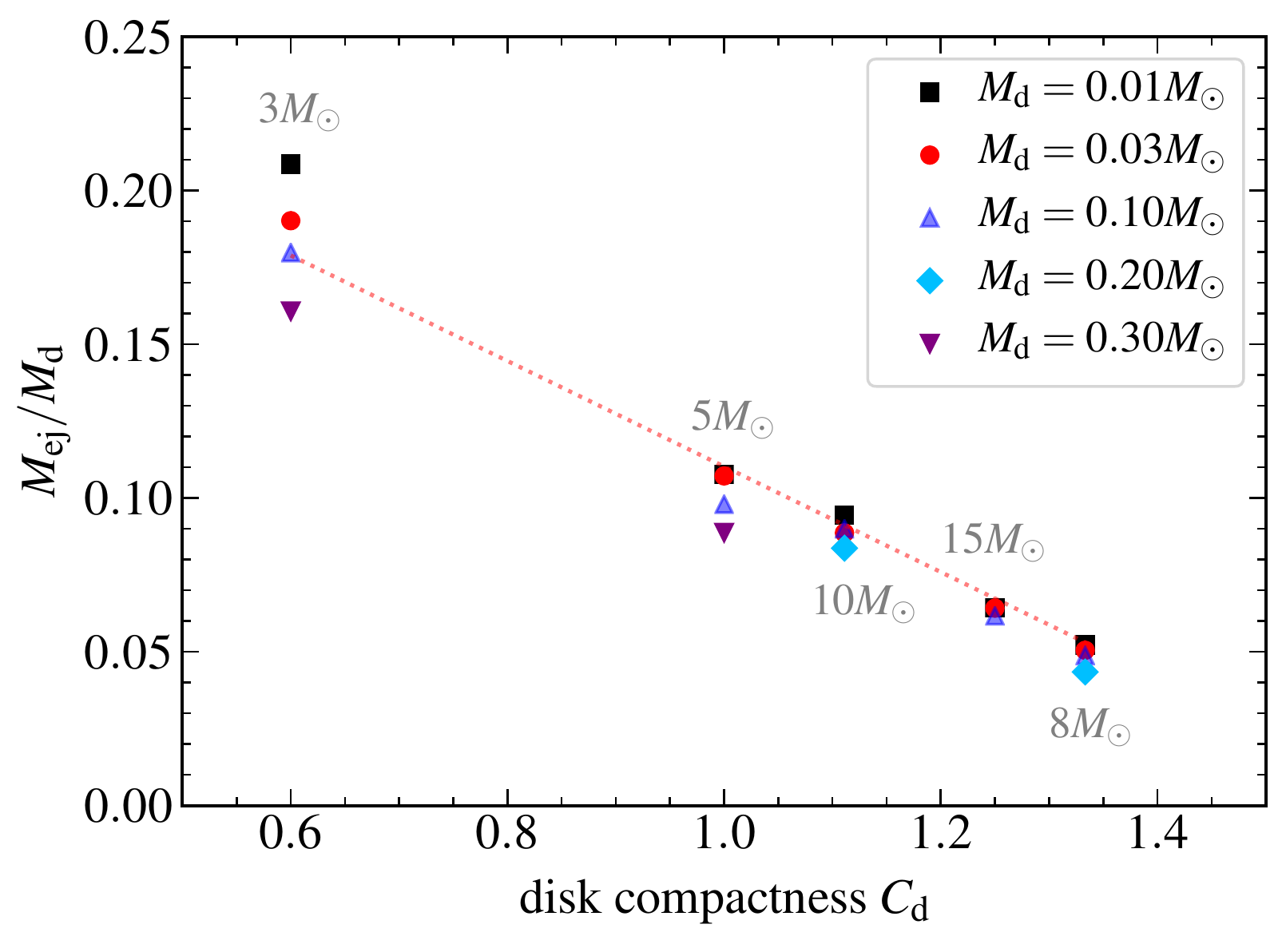}
\caption{\emph{Left:} Fraction of the initial disk mass ejected with positive 
Bernoulli parameter (eq.~\ref{eq:berc_general})
as a function of BH mass. Different symbols and colors correspond to different
disk masses, as labeled.  The gray number above each symbol column corresponds
to the disk compactness parameter $C_{\rm d}$
(eq.~\ref{eq:disk_compactness_def}).  \emph{Right:} Fraction of the initial
disk mass ejected as a function of disk compactness parameter, using the same
color and symbol coding as in the left panel. The gray numbers above each
symbol column denote the corresponding BH mass. The red dotted line is a linear
fit to the ejected fraction for disks with $M_{\rm d}=0.03M_\odot$. GRMHD effects
can enhance the ejected fraction and average velocity by up to 
a factor of $\sim 2$ relative to hydrodynamic models \citep{F19_grmhd}.}
\label{f:mass_ejection}
\end{figure*}

%-----------------------------------------------------------------------------------
\subsection{Mass ejection}
\label{s:mass_ejection}

The disk mass ejection rate in all directions $\dot{M}_{\rm out}$ is measured at 
a radius $r_{\rm out}=10^9$\,cm, far enough outside
the disk that mostly complete ejection is achieved before the disk viscously spreads to
that radius. Material is considered to be unbound when its Bernoulli parameter
\begin{equation}
\label{eq:berc_general}
b = \frac{1}{2}\left[v_r^2 + v_\theta^2 + \left(\frac{j}{r\sin\theta}\right)^2 \right] + e_{\rm int} + \frac{p}{\rho} + \Phi
\end{equation}
is positive. In equation~(\ref{eq:berc_general}), $v_r$ and $v_\theta$ are the radial and meridional velocities,
$e_{\rm int}$ is the specific internal energy, $p$ is the pressure, $\rho$ is the density, $j$ is the
specific angular momentum, and $\Phi$ is the gravitational potential.  
Table~\ref{t:results} shows the ejected outflow mass $M_{\rm ej}$ 
-- the time integral of $\dot{M}_{\rm out}$ over the simulation --
for all models. 

The fraction of the initial disk mass ejected is shown in Figure~\ref{f:mass_ejection} for the 
baseline model sequence. This fraction ranges from $4\%$ for the heaviest disk with 
$M_{\rm bh}=8M_\odot$ to $21\%$ for the lightest disk around the $M_{\rm bh}=3M_\odot$ BH. The most
important trend in mass ejection apparent from Table~\ref{t:results} is a monotonically decreasing ejected
fraction with strength of gravity at the disk, $\propto M_{\rm bh}/R_{\rm d}$. 
For convenience, we define a disk compactness parameter as
\begin{equation}
\label{eq:disk_compactness_def}
C_{\rm d} = \left(\frac{M_{\rm bh}}{5M_\odot}\right)\left(\frac{50\,\textrm{km}}{R_{\rm d}}\right).
\end{equation}
Figure~\ref{f:mass_ejection} also shows that 
the dependence of the ejected fraction with compactness is approximately linear, although
at low BH masses the disk mass becomes an additional factor. This dependence on the
strength of gravity has previously been documented in \citet{FM13}, and is also apparent from the results
of \citet{just2015}, \citet{fernandez_2017}, and \citet{fujibayashi2020}, although this is the first 
time that it is sampled over an extended region of parameter space. 

The second mass ejection trend in all models of the baseline sequence is a
monotonic decrease in the ejected fraction with increasing disk mass at constant compactness. 
Figure~\ref{f:mass_ejection}
shows that the strength of this dependence on disk mass is itself a function of disk compactness, 
with low-compactness disks being the most sensitive to the initial disk mass, while in high compactness
systems this property has a smaller impact on the ejected fraction.

\begin{figure}
\includegraphics*[width=\columnwidth]{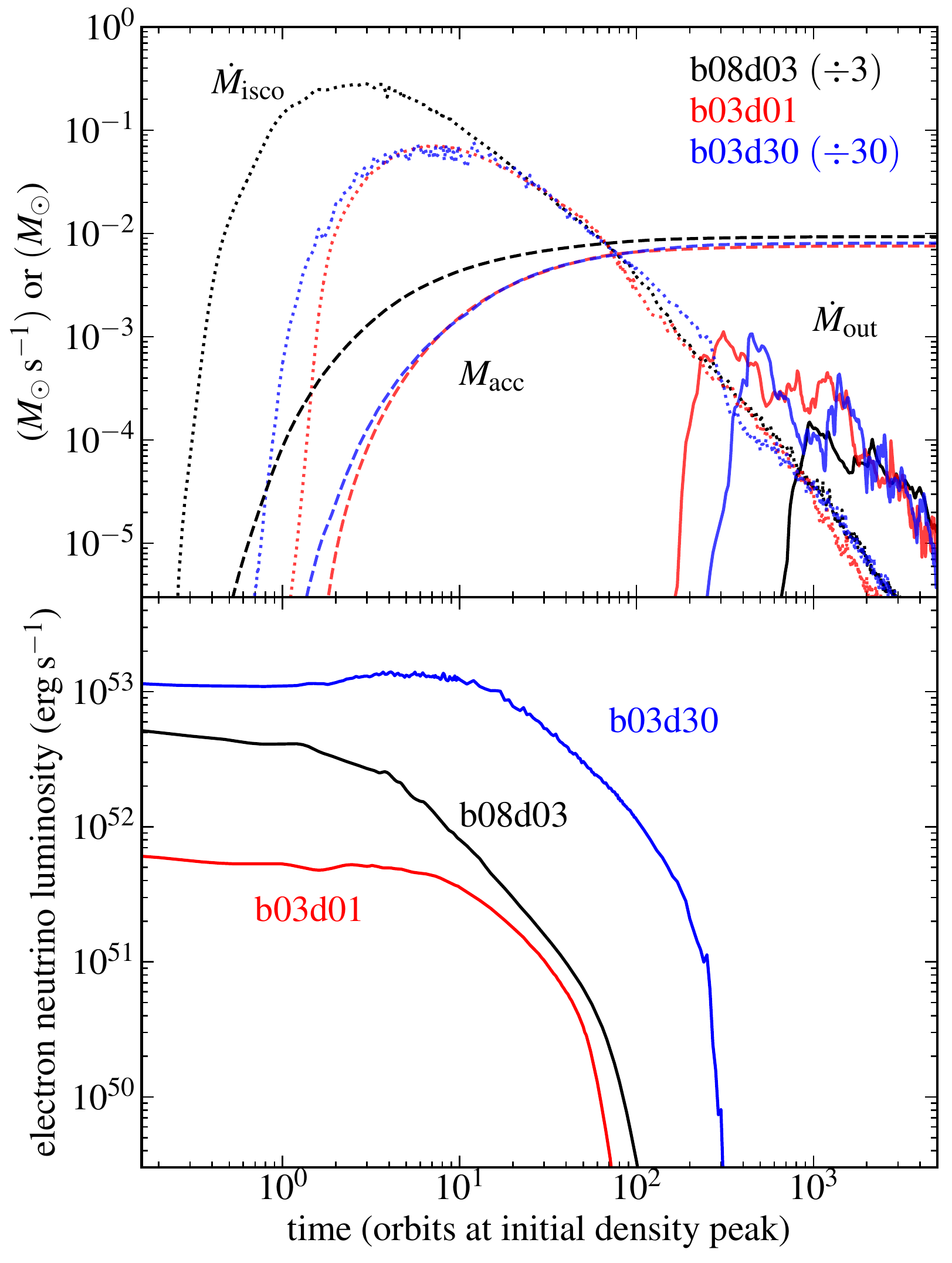}
\caption{\emph{Top:} Mass accretion rate at the ISCO ($\dot M_{\rm isco}$, dotted lines), cumulative accreted
mass at the ISCO ($M_{\rm acc}$, dashes lines), and mass outflow rate in unbound material at $r=10^9$\,cm 
($\dot M_{\rm out}$, solid lines) as a function of time for models b08d03 (high-compactness), 
b03d01 (low compactness, low disk mass), and b03d30 (low compactness, high disk mass), as labeled. 
To facilitate comparison, the data from models b08d03 and b03d30 has been normalized to a disk
mass of $0.01M_\odot$ (as in model b03d01). \emph{Bottom:} Electron neutrino luminosity
for models b08d03, b03d01, and b03d30, as labeled. The fraction of the disk ejected is
related to the fraction of the disk accreted at the time when weak interactions
freeze out.}
\label{f:mej_lum_time}
\end{figure}

The physical origin of these trends in mass ejection can be traced back to the nature of the
ejection mechanism. Most of the outflow is launched once weak interactions freeze
out in the disk, removing the source of cooling. The fraction of the disk mass available to be ejected
depends on how much has already been lost to accretion onto the BH by the time 
freezout occurs. This interplay is illustrated in Figure~\ref{f:mej_lum_time}, which shows the evolution of the
mass accretion rate ($\dot{M}_{\rm isco}$) and cumulative mass accreted at the ISCO ($M_{\rm acc}$), 
mass ejection rate at large radii ($\dot{M}_{\rm out}$), and the electron neutrino luminosity. 

In the model with the highest compactness (b08d03), accretion starts much earlier when measured in orbital
times than in the lower-compactness models. By the time weak interactions freeze out (steep plummet in 
neutrino luminosity at about $100$ orbits) a significant fraction of the disk ($85\%$) has already been 
accreted to the BH. This earlier onset of accretion, despite having the same viscosity parameter, is due to the disk
being closer to the ISCO. In terms of dimensionless numbers: $r_{\rm isco}/R_{\rm d}=\{0.26,0.57\}$ 
for $C_{\rm d}=\{0.60,1.33\}$, respectively.

For disks of the same compactness, the evolution of the accretion rate is very similar.
Figure~\ref{f:mej_lum_time} shows that mass ejection begins later in the more massive
disk, which also takes longer time to reach freezout of weak interactions. 
More massive disks are more optically thick to neutrinos owing to their higher
initial density, and for the same strength of viscosity, it takes more orbits for 
the density to decrease to a level where neutrino processes are no longer effective in
cooling the disk. At the time when the neutrino luminosity reaches $10^{50}$\,erg\,s$^{-1}$,
models b03d01 and b03d30 have accreted $60\%$ and $78\%$ of their initial disk masses, respectively.

\begin{figure}
\includegraphics*[width=\columnwidth]{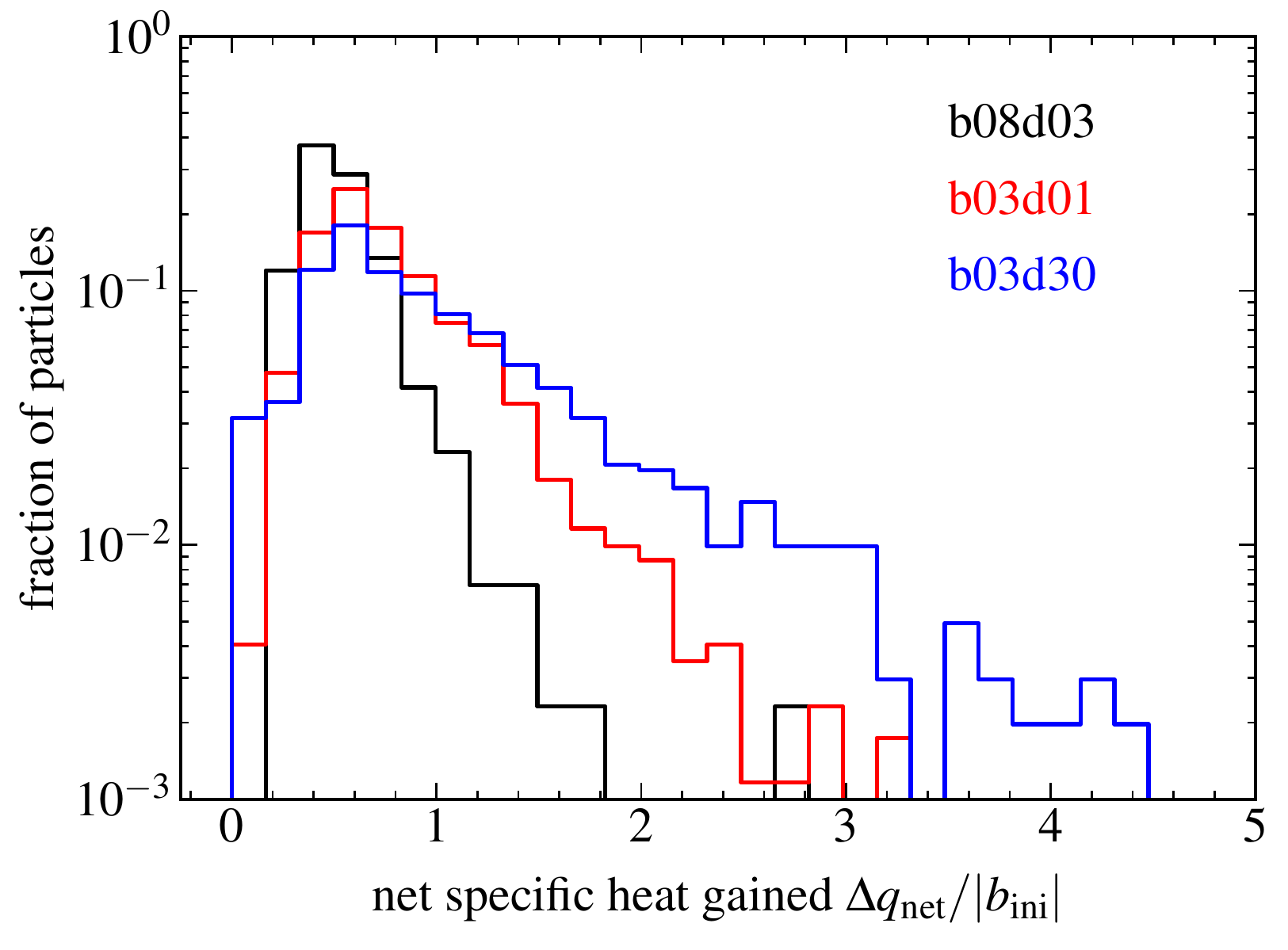}
\caption{Distribution of net specific heat gained by tracer particles due to source terms 
(eq.~\ref{eq:dq_net}) in the hydrodynamic evolution of models b08d03, b03d01, and b03d30, as labeled.
The heat gain is normalized to the initial Bernoulli parameter in the disk
(eq.~\ref{eq:berc_disk}).}
\label{f:hist_enet_differences}
\end{figure}

\begin{table*}
\begin{minipage}{17cm}
\centering
\caption{Average time-integrated heat gain or loss per unit mass (eqns.~\ref{eq:dq_visc}-\ref{eq:dq_net}), 
and change in $Y_e$ (eqns.~\ref{eq:gcem}-\ref{eq:dye_definition}), due to various processes acting
on tracer particles during the hydrodynamic evolution of selected models. The specific heat gain due to
process $i$ is normalized as $\bar{\Delta} q_{i,19} = \bar{\Delta} q_i/(10^{19}\,\textrm{erg\,g}^{-1})$,
and $\bar{\Delta}$ denotes average over all outflow particles. Each quantity is separately
rounded for clarity, net sums match when including all significant digits.}
\label{t:trajectory_integrals}
\begin{tabular}{lcccccccccccc}
\hline
Model & $\bar\Delta q_{\rm net}/|b_{\rm ini}|$ & $\bar\Delta q_{\rm net,19}$ & $\bar\Delta q_{\rm visc,19}$ & 
        $\bar\Delta q_{\nu,19}$ & $\bar\Delta q_{\alpha,19}$ &
        $\bar\Delta Y_e^{\rm net} $ & $\bar\Delta Y_e^{\rm em,\nu_e}$ & $\bar\Delta Y_e^{\rm em,\bar{\nu}_e}$ & 
        $\bar\Delta Y_e^{\rm abs,\nu_e}$ & $\bar\Delta Y_e^{\rm abs,\bar{\nu}_e}$ & 
        $\bar\Delta Y_e^{\rm em}$ & $\bar\Delta Y_e^{\rm abs}$\\
\hline
b03d01     & 0.81 & 2.1 & 2.6 & -0.8 & 0.3 & 0.08 & 0.42 & 0.45 & 0.07 & 0.01 &  0.03 & 0.05 \\
b03d03     & 0.79 & 1.7 & 2.5 & -1.1 & 0.3 & 0.08 & 0.64 & 0.61 & 0.15 & 0.04 & -0.03 & 0.12 \\
b03d10     & 0.91 & 1.5 & 2.8 & -1.6 & 0.3 & 0.09 & 1.00 & 0.91 & 0.25 & 0.07 & -0.09 & 0.18 \\
b03d30     & 1.12 & 1.3 & 2.9 & -1.9 & 0.3 & 0.09 & 1.12 & 1.04 & 0.26 & 0.08 & -0.08 & 0.18 \\
\noalign{\smallskip}
b08d01     & 0.51 & 3.5 & 3.9 & -0.7 & 0.3 & 0.09 & 0.34 & 0.41 & 0.03 & 0.01 &  0.07 & 0.02\\
b08d03     & 0.55 & 3.4 & 4.2 & -1.1 & 0.4 & 0.10 & 0.53 & 0.58 & 0.07 & 0.02 &  0.05 & 0.05\\
b08d10     & 0.60 & 3.2 & 4.5 & -1.7 & 0.4 & 0.11 & 0.81 & 0.83 & 0.13 & 0.04 &  0.02 & 0.09\\
b08d20     & 0.66 & 3.1 & 4.9 & -2.2 & 0.4 & 0.13 & 1.10 & 1.08 & 0.21 & 0.06 & -0.02 & 0.15\\
\noalign{\smallskip}
\noalign{\smallskip}
b03d01-y10 & 0.79 & 2.1 & 2.6 & -0.8 & 0.3 & 0.16 & 0.37 & 0.48 & 0.07 & 0.01 & 0.11 & 0.05\\
b03d01-y15 & 0.78 & 2.1 & 2.6 & -0.8 & 0.3 & 0.12 & 0.40 & 0.47 & 0.07 & 0.01 & 0.07 & 0.05\\
b03d01-v10 & 0.94 & 2.5 & 2.7 & -0.5 & 0.3 & 0.07 & 0.22 & 0.25 & 0.05 & 0.02 & 0.04 & 0.03\\
\noalign{\smallskip}
b03d30-y10 & 1.16 & 1.4 & 3.1 & -2.1 & 0.3 & 0.19 & 1.16 & 1.16 & 0.27 & 0.08 &  0.00 & 0.19\\
b03d30-y15 & 1.05 & 1.3 & 2.7 & -1.8 & 0.3 & 0.13 & 1.05 & 1.01 & 0.25 & 0.07 & -0.04 & 0.17\\
\noalign{\smallskip}
b08d03-y10 & 0.53 & 3.3 & 4.0 & -1.1 & 0.3 & 0.19 & 0.46 & 0.60 & 0.07 & 0.02 & 0.14 & 0.05 \\
b08d03-y15 & 0.55 & 3.4 & 4.3 & -1.2 & 0.3 & 0.15 & 0.55 & 0.65 & 0.07 & 0.02 & 0.10 & 0.05 \\
b08d03-v10 & 0.67 & 4.2 & 4.9 & -1.0 & 0.4 & 0.12 & 0.40 & 0.48 & 0.06 & 0.02 & 0.08 & 0.04 \\
\hline
\end{tabular}
\end{minipage}
\end{table*}

The density dependence of the neutrino optical depth is not the only factor influencing the
freezout time. Table~\ref{t:models} and equation~(\ref{eq:berc_disk}) show that to keep the entropy
constant, initial equilibrium disks with higher masses also have a higher internal energy 
content and therefore higher temperatures. This effect is stronger 
for lower compactness configurations. While disk material is more weakly bound, it can also radiate neutrinos
at relevant levels for a longer time, and so this acts in the direction of delaying 
the onset of mass ejection. The post-merger entropy of the disk is thus an important parameter
to keep track of in dynamical merger simulations given its effect on mass ejection efficiency.

Using the outflow trajectories from hydrodynamic simulations we can further
analyze the energetics of mass ejection. In particular, we can quantify the strength
and importance of different processes that change the heat content of the fluid:
viscous heating, neutrino heating/cooling, and nuclear recombination of alpha particles.
For each outflow trajectory, we compute the time integral of the local energy source term, 
yielding the heat gained or lost by the fluid element per unit mass:
\begin{eqnarray}
\label{eq:dq_visc}
\Delta q_{\rm visc} & = & \int_0^{t_{\rm max,p}} \dot{q}_{\rm visc} dt\\
\label{eq:dq_nu}
\Delta q_\nu        & = & \int_0^{t_{\rm max,p}} \dot{q}_{\rm \nu} dt\\
\label{eq:dq_alpha}
\Delta q_\alpha     & = & \frac{B_\alpha}{m_\alpha}\left [X_{\rm alpha}(t_{\rm max,p}) - X_\alpha(0) \right]\\
\label{eq:dq_net}
\Delta q_{\rm net}  & = & \Delta q_{\rm visc} + \Delta q_\nu + \Delta q_\alpha,
\end{eqnarray}
where $\{\dot{q}_{\rm visc},\dot{q}_{\rm \nu}\}$ stand for viscous heating and neutrino heating/cooling, 
respectively, ${B_\alpha}/{m_\alpha}\simeq 6.8\times 10^{18}$\,erg\,g$^{-1}$ is the specific nuclear 
binding energy of alpha particles, $X_\alpha$ is the mass fraction of alpha particles, 
$t_{\rm max,p}$ is the maximum time for the particle evolution, either when it leaves the outer boundary of the 
computational domain or when the simulation ends, whichever is shorter, and $\Delta q_i$ is the heat gained or 
lost from process $i$. The resulting quantities are shown in Table~\ref{t:trajectory_integrals} 
for selected models.

\begin{figure}
\includegraphics*[width=\columnwidth]{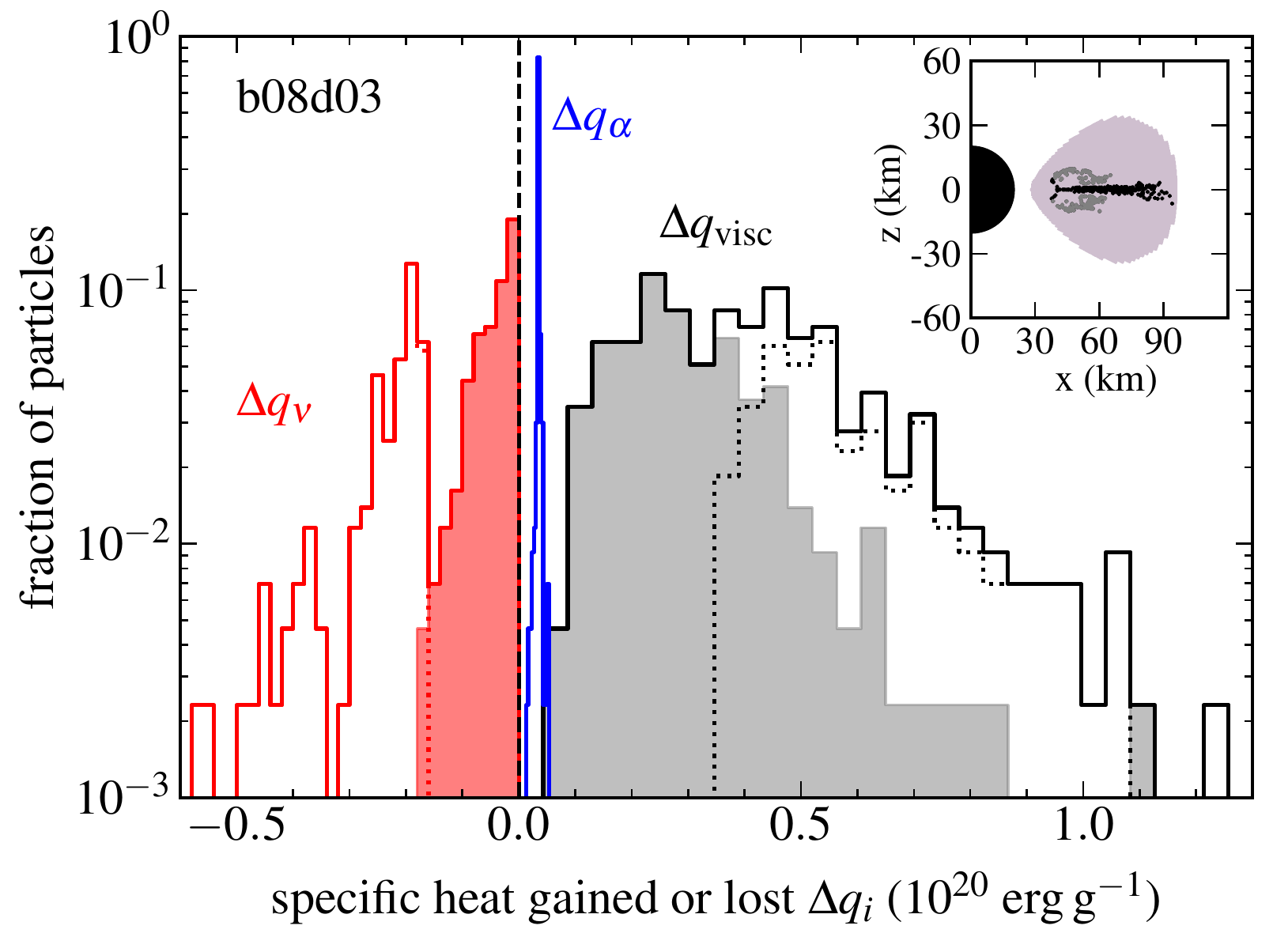}
\caption{Distribution of specific heat gained or lost by tracer particles due to source terms 
(eqns.~\ref{eq:dq_visc}-\ref{eq:dq_alpha}), as labeled, in the hydrodynamic evolution
of model b08d03. The inset shows a snapshot of the
initial positions of outflow particles in the disk, with gray/black particles corresponding
to the shaded/dotted subsets of the neutrino and viscous heat gain histograms, 
respectively (see also \citealt{wu_2016} for 
a larger plot of the initial particle distribution).}
\label{f:heat_breakdown}
\end{figure}

Figure~\ref{f:hist_enet_differences} shows the distribution of net heat gained 
$\Delta q_{\rm net}$ (eq.~\ref{eq:dq_net}) 
by outflow particles
for the same set of simulations shown in Figure~\ref{f:mej_lum_time}.
Specific energies are shown in units of the initial Bernoulli parameter of the disk (equation~\ref{eq:berc_disk}).
In all three models, the distribution shows a peak at around $60\%$ of $|b_{\rm ini}|$, with
a tail toward high gain that is more extended for less compact models and higher disk masses.
For a large fraction of the outflow, unbinding is not completely achieved by absorption of
heat alone, which means that a significant part of the energy gain comes from adiabatic work.
Disks with higher compactness are also less effective at absorbing heat, despite the
fact that the absolute value of the net heat gain is higher (Table~\ref{t:trajectory_integrals}).

It is worth emphasizing that while disks with a higher compactness are more gravitationally
bound in an absolute sense ($b_{\rm ini} \propto M_{\rm bh}/R_{\rm d}$ in eq.~\ref{eq:berc_disk}), they also undergo
more net heating than lower compactness models (Table~\ref{t:trajectory_integrals}). 
Thus the absolute value of the gravitational
potential (in a Newtonian sense), while correlating negatively with the fractional amount of heat absorbed,
does not by itself explain the efficiency of mass ejection without accounting for how
close the disk is to the ISCO radius.

For individual models, viscous heating and nuclear recombination contribute with net heating,
while neutrinos primarily cool the disk (Table~\ref{t:trajectory_integrals}). 
This is illustrated in Figure~\ref{f:heat_breakdown},
which shows the distribution of the individual heating/cooling terms for the outflow
from model b08d03. While there is a non-negligible fraction of particles for which neutrinos
provide net heating, the magnitude of this heating ($\sim 10^{17}$\,erg\,g$^{-1}$) is
dynamically negligible when compared to the dominant source terms ($\sim 10^{19}$\,erg\,g$^{-1}$).

The integrated heat gain due to $\alpha$ particle recombination in the hydrodynamic simulation
is sub-dominant compared to that from 
viscous heating. Despite its low global value, however, the heating due
to $\alpha$ recombination is deposited over a short amount of time as the outflow is launched, and
it can become comparable or even exceed the rate of viscous heating in this phase. The bulk of
neutrino heating and cooling takes place before weak freezout. 
Figure~\ref{f:heat_breakdown} shows that the nuclear recombination gain is narrowly distributed
around the mean value, $3-4\times 10^{18}$\,erg\,g$^{-1}$ (Table~\ref{t:trajectory_integrals}).
This value can be understood from the fact that upon expansion and cooling, all fluid elements achieve 
the maximal alpha particle mass fraction set by charge conservation, $X_{\alpha,\rm max} = 2Y_e$.
The average electron fraction of the outflow $\langle Y_e\rangle\simeq 0.3$ (Table~\ref{t:results}), 
then sets the average amount of energy gained. 

Figure~\ref{f:heat_breakdown} also shows that the tracer particles in model b08d03 follow bimodal
distributions of viscous heating and neutrino cooling. This bimodality can be traced back to
the initial positions of the particles in the disk. These particles originate from two regions: (1)
the equatorial plane of the disk, and (2) regions above the equatorial plane around the initial
density peak (see also \citealt{wu_2016}). The first group of particles experiences stronger
viscous heating and neutrino cooling, while the second group experiences heating or cooling
with less intensity. The latter group includes all the particles that experience net neutrino
heating. The initial position of the particles is related to the way the disk overturns
in the poloidal direction due to viscous heating, and may differ from that obtained when
MHD turbulence transports angular momentum.

The fraction of the disk mass ejected and average outflow velocity are nearly insensitive to 
the initial $Y_e$ of the disk except for very massive disks in low-compactness systems,
where differences of a few percent of the disk mass can arise (Table~\ref{t:trajectory_integrals}). 
Changes in the viscosity parameter, on the other hand, result in important changes to both 
ejected fraction and outflow velocity. 
The average outflow radial velocity is in the range  $0.03-0.04$\,c for all models that use 
$\alpha=0.03$, while this average increases to $0.05-0.06$\,c for the models with $\alpha=0.1$. 
The fraction of the disk mass ejected increases by $\sim 30\%$ for the model with low compactness 
(b03d01-v10) and by a factor of $2$ for the high compactness model (b08d03-v10).
Table~\ref{t:trajectory_integrals} shows that the increase in the viscosity parameter results in 
more viscous heating in the high-compactness model b08d03-v10 and less neutrino cooling
in the least compact model b03d01-v10, in both cases increasing the net
heat gain of the outflow.

\begin{figure}
\includegraphics*[width=\columnwidth]{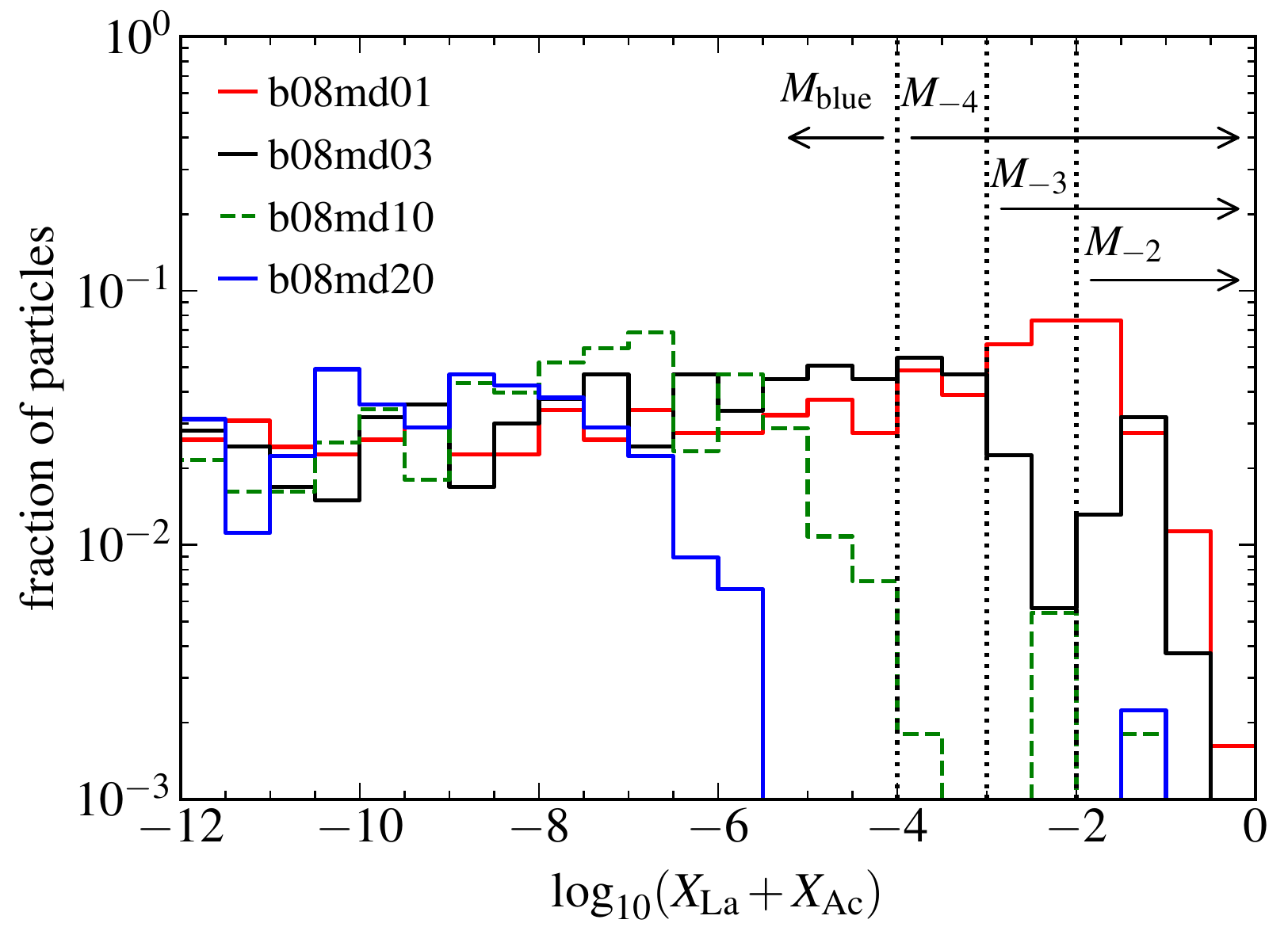}
\includegraphics*[width=\columnwidth]{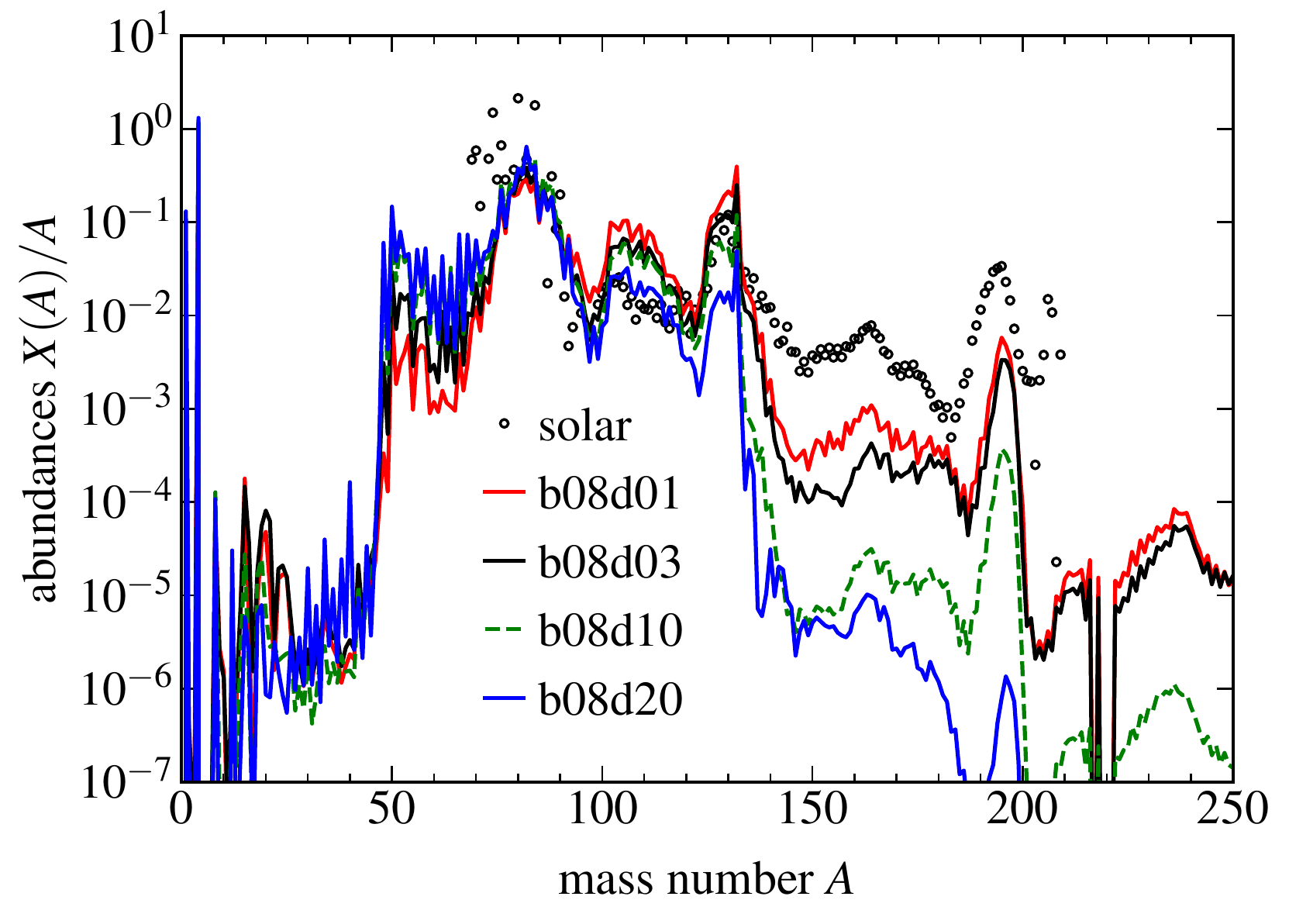}
\caption{\emph{Top:} Distribution of lanthanide and actinide mass fractions at 1 day, for
nuclear-network-processed particles from models b08d01-b08d20, as labeled.
Since each particle represents an equal mass element in the disk, a sum over the bins yields the
fraction of the mass with a given Lanthanide and Actinide fraction ($M_{\rm blue}$ and
$\{M_{-4},M_{\-3},M_{-2}\}$, c.f. eqns.~(\ref{eq:Mblue_def})-(\ref{eq:Mred_def}) and 
Table~\ref{t:results}). The histograms continue to mass fractions 
lower than $10^{-12}$ with similar slope,
and were truncated for clarity. \emph{Bottom:} Isotopic abundances at 1 day for nuclear-network-processed 
particles from models b08d01-b08d20, as labeled. Abundances are normalized such that their mass
fractions add up to unity. The solar system $r$-process abundances from \citet{goriely1999} are normalized
to model b08d03 at $A=130$. Note that the dynamical ejecta is rich in elements with $A>130$, and
in combination with the disk outflow can supply the entire range of $r$-process elements (e.g., \citealt{just2015}).}
\label{f:hist_lan+ac_abundances}
\end{figure}

%-----------------------------------------------------------------------------------
\subsection{Outflow Composition}

Table~\ref{t:results} shows that the average electron fraction of the disk outflow is in
the range $0.25-0.35$ for all simulated models. These values are just above the critical 
transition at which the nucleosynthesis changes from rich to poor in elements
with mass numbers $A>130$ (e.g., \citealt{kasen_2015,lippuner2015}). For the purposes of predicting
kilonova properties, the mass fraction in lanthanides ($57\leq Z \leq 72$, with $Z$ the atomic number) 
and actinides ($89\leq Z\leq 104$) is the most
important, since these species have an outsize influence on the ejecta opacity
-- and therefore on the kilonova color, luminosity, and duration -- given their atomic 
complexity \citep{tanaka2013,Kasen+13,Barnes&Kasen13,fontes2015,tanaka_2019}.

We therefore refine our diagnostic of the outflow composition by analyzing the output 
of post-processed tracer particles with {\tt SkyNet} (\S\ref{s:skynet}). For
each simulation, we report in Table~\ref{t:results} the fraction of the outflow particles with
lanthanide and actinide mass fraction $X_{\rm La}+X_{\rm Ac}< 10^{-4}$ and define it
as the `blue mass'
\begin{equation}
\label{eq:Mblue_def}
M_{\rm blue} = \int_{0}^{10^{-4}}\frac{dM_{\rm ej}}{d(X_{\rm La}+X_{\rm Ac})}\,d(X_{\rm La}+X_{\rm Ac}).
\end{equation} 
The value of $10^{-4}$ is small enough that the outflow opacity is indistinguishable
from that dominated by iron-group elements \citep{Kasen+13}. Likewise, we define three `red' masses
\begin{equation}
\label{eq:Mred_def}
M_{-k} = \int_{10^{-k}}^1 \frac{dM_{\rm ej}}{d(X_{\rm La}+X_{\rm Ac})}\,d(X_{\rm La}+X_{\rm Ac}).
\end{equation}
using $-k=\{-4,-3,-2\}$. These four numbers provide a description of the incidence of
heavy $r$-process elements in the ejecta that, while coarser than detailed abundances,
is more informative than just the electron fraction. 
Figure~\ref{f:hist_lan+ac_abundances} shows the distribution of mass fractions at 1 day
in the outflow particles from models b08d01-b08d20 (highest compactness), along with
the regions encompassed by $M_{\rm blue}$ and $M_{-k}$, and the isotopic abundances for comparison.
At 1 day, the mass fractions of actinides are in general $10$ times smaller than that
of lanthanides in our models, and thus we lump them together when discussing composition
effects (for more detailed studies on actinide production see \citealt{eichler_2019} or \citealt{holmbeck_2019}).

The most robust composition trends from Table~\ref{t:results} are that (1) the majority
of the disk outflow mass is lanthanide poor ($M_{\rm blue}$), and that (2)
the fraction of lanthanide poor material is a monotonically increasing function of the 
disk mass, for constant BH mass. The dependence on disk mass is clearly illustrated in 
Figure~\ref{f:hist_lan+ac_abundances}: the fraction of particles with high lanthanide
abundance is a steep function of the disk mass, as is the abundance of of
elements with $A > 130$ (this dependence has also been reported by 
\citealt{just2015} and \citealt{fujibayashi2020}). While there is
some dependence on the compactness $C_{\rm d}$ at constant disk mass, this dependence is not
fully monotonic, and is weaker than the dependence on disk mass at fixed compactness. Therefore,
this variation with compactness is more likely to be dependent on the details
of how the disk evolution is modeled. 

\begin{figure}
\includegraphics*[width=\columnwidth]{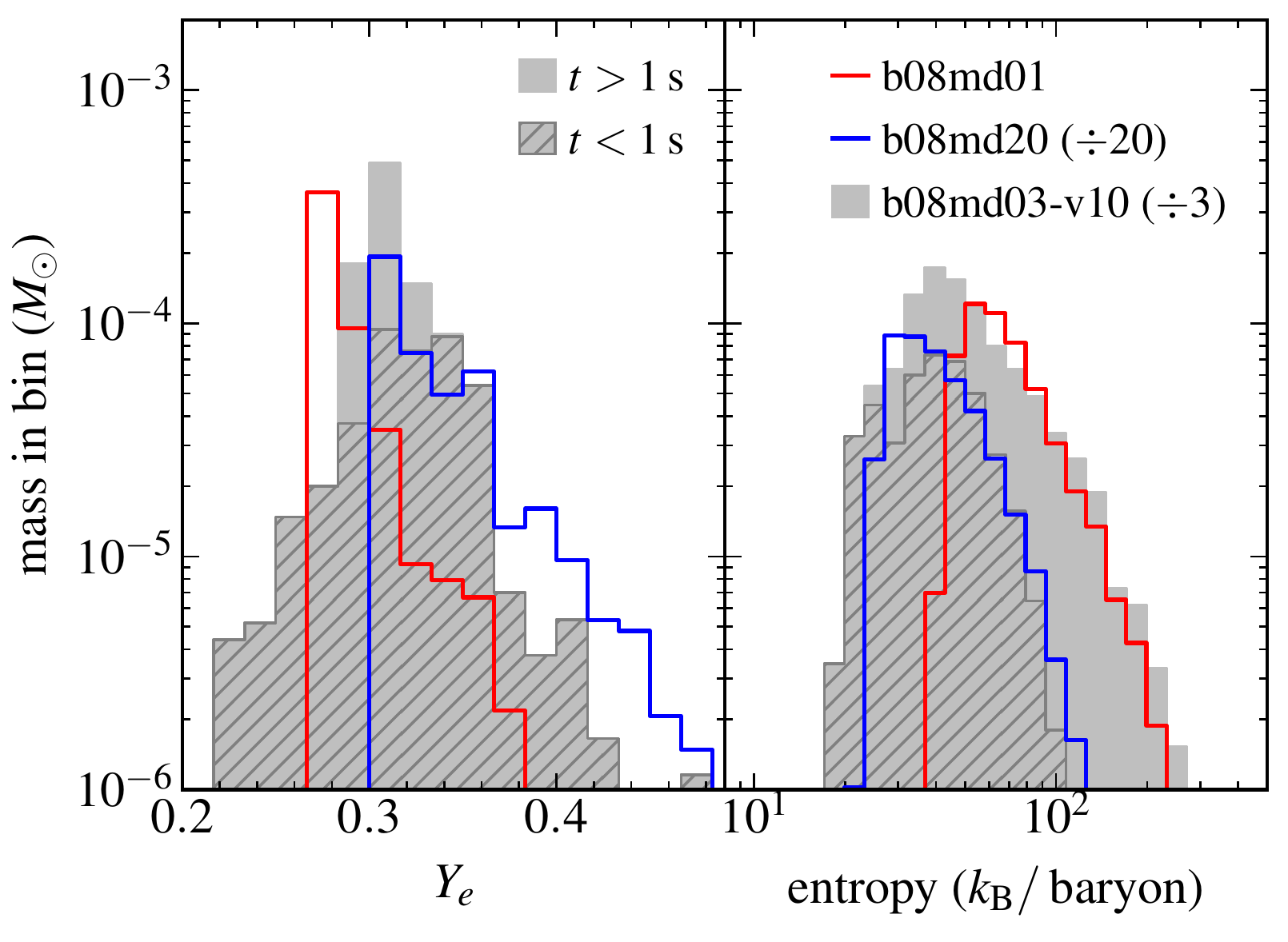}
\caption{Mass histograms of electron fraction (left) and entropy (right) for models
b08d01 (low disk mass), b08d20 (high disk mass), and b08d03-v10 (high viscosity), obtained
by the end of the hydrodynamic simulation at $r=10^{9}$\,cm. To facilitate comparison, 
histogram masses have been normalized to that of model b08d01. The hatched region is the
subset of the histogram of model b08d03-v10 for times $t<1$\,s .}
\label{f:hist_ye_entropy_b08}
\end{figure}

The trend of more lanthanide poor ejecta with increasing disk mass is apparent
from the distribution of electron fraction (Figure~\ref{f:hist_ye_entropy_b08}). 
Despite having a very similar average value, the $Y_e$ distribution of model 
b08d20 extends to significantly higher values than in model b08d01. Naively,
one would expect that higher electron fraction is associated with a higher
equilibrium $Y_e$ arising from a higher abundance of positrons given higher 
entropies (e.g., \citealt{Beloborodov03}). However, the entropy distributions
of models b08d01 and b08d20 show that the average entropy \emph{decreases} with
higher disk masses. The lower entropy can be understood from the fact that the temperature is
primarily set by the strength of the gravitational potential once accretion is established, 
and is similar in both models. On the other hand, the densities are higher at any given 
time for a higher disk mass, with a correspondingly lower entropy than at lower disk masses.

\begin{figure}
\includegraphics*[width=\columnwidth]{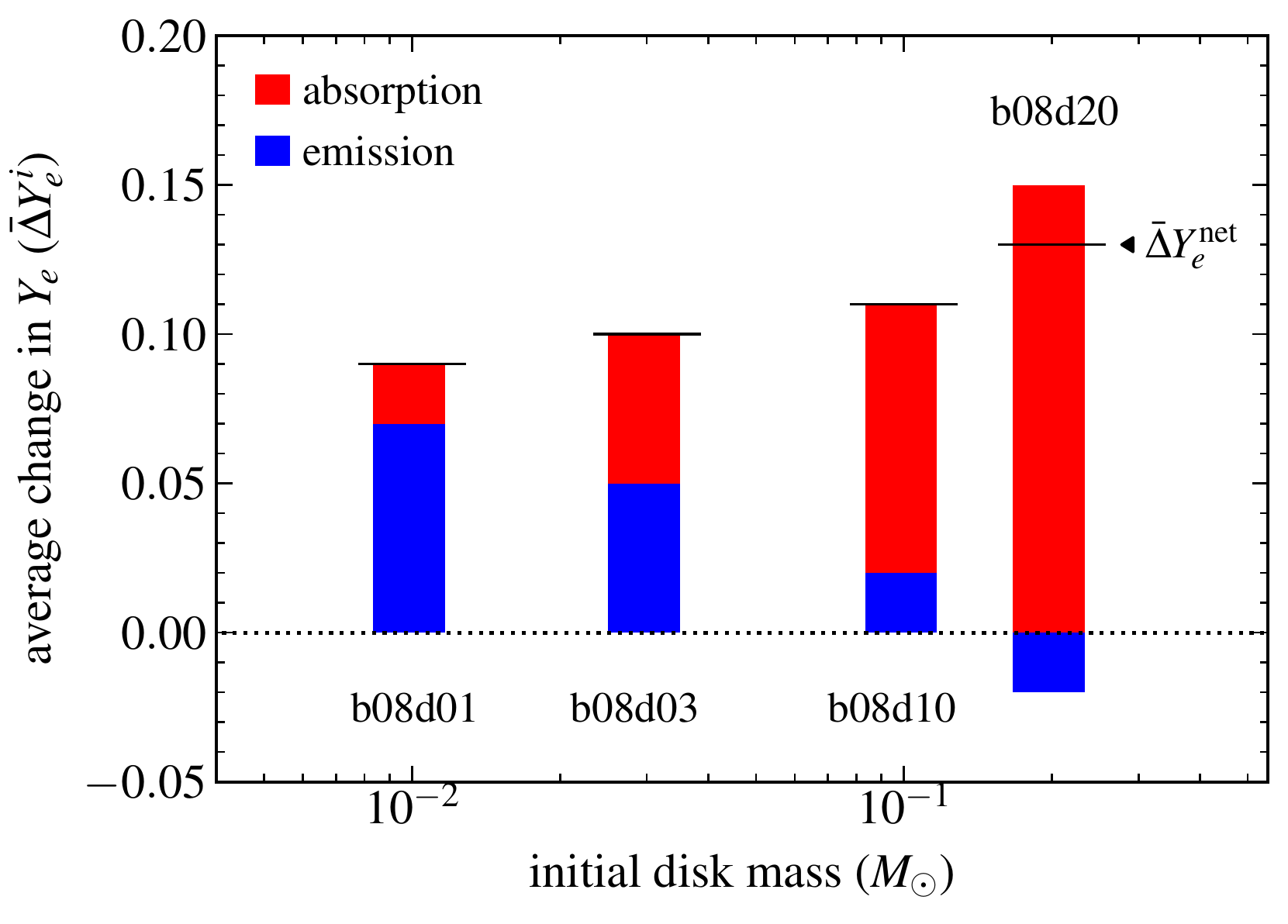}
\caption{Average electron fraction change in particles due to neutrino emission or absorption 
(eqns.~\ref{eq:gcem}-\ref{eq:gamn} and \ref{eq:dye_definition}, Table~\ref{t:trajectory_integrals}), 
during the hydrodynamic evolution of models b08d01-b08d20, as labeled. The black 
horizontal lines denote the net change in $Y_e$ when including all processes. The relative importance
of neutrino absorption increases as the disk mass increases, all else being equal, accounting
for the higher $Y_e$ of the outflow with lower overall entropies (Figure~\ref{f:hist_ye_entropy_b08}).}
\label{f:delta-ye_mdisk}
\end{figure}

We can further analyze the evolution of the electron fraction distribution 
of the outflow by computing the time-integrated contribution of the different 
neutrino processes that change $Y_e$. The net rate of change of $Y_e$ due to 
neutrino emission and absorption arises from the following reactions:
\begin{eqnarray}
\label{eq:gcem}
\Gamma^{\rm em, \nu_e}:        && e^- + p \to n + \nu_e\\
\label{eq:gcep}
\Gamma^{\rm em, \bar{\nu}_e}:  && e^+ + n \to p + \bar{\nu}_e\\
\label{eq:ghem}
\Gamma^{\rm abs, \nu_e}:       && \nu_e + n \to  e^- + p\\
\label{eq:ghep}
\Gamma^{\rm abs, \bar{\nu}_e}: && \bar{\nu}_e + p \to  e^+ + n\\
\noalign{\smallskip}
\Gamma^{\rm em}   & = &  \Gamma^{\rm em, \bar{\nu}_e} - \Gamma^{\rm em, \nu_e}\\
\Gamma^{\rm abs} & = & \Gamma^{\rm abs, \nu_e} - \Gamma^{\rm abs, \bar{\nu}_e}\\
\label{eq:gamn}
\Gamma^{\rm net} & = & \Gamma^{\rm em} + \Gamma^{\rm abs}
\end{eqnarray}
with the net rate setting the overall evolution of the electron fraction in the
hydrodynamic simulation: 
$\partial Y_e/\partial t + \mathbf{v}\cdot \nabla Y_e = \Gamma^{\rm net}$.
For each tracer particle, we compute a separate time integral for each of
the rates above, obtaining a contribution to the change in electron fraction:
\begin{equation}
\label{eq:dye_definition}
\Delta Y_e^i = \int_0^{t_{\rm max,p}} \Gamma^i dt,
\end{equation}
where again $t_{\rm max,p}$ is the maximum time of the particle in the simulation.

Table~\ref{t:trajectory_integrals} shows that the rates of neutrino emission dominate
over neutrino absorption for all models, in line with the overall dominance of neutrino 
cooling over neutrino heating (c.f. Figure~\ref{f:heat_breakdown}). 
However, the change in $Y_e$ is set by \emph{differences}
in the rates of neutrino/antineutrino emission and absorption. If the two emission rates 
are closer in magnitude than the two absorption rates, the latter can dominate the change 
in $Y_e$ despite being smaller in magnitude than the former.

Figure~\ref{f:delta-ye_mdisk} shows the average change in $Y_e$ for tracer particles 
as a function of disk mass in the baseline sequence with $M_{\rm bh}=8M_\odot$, 
along with the breakdown of this change between emission and absorption of electron 
neutrinos/antineutrinos. At low disk masses, emission processes dominate the change
in $Y_e$, with decreasing relative importance with increasing
disk mass. Emission processes act toward driving $Y_e$ to its local equilibrium value set 
by the entropy, and this equilibrium is lower at higher disk masses given the lower entropy
(Figure~\ref{f:hist_ye_entropy_b08}). At the highest disk mass (model b08d20),
this change in $Y_e$ due to emission is even negative. 

Absorption, on the other hand, is set by the ambient flux of incident 
neutrinos and the mass fractions of neutrons and protons. At higher disk masses,
neutrino/antineutrino luminosities are higher and stay high for a longer time 
(Figure~\ref{f:mej_lum_time}) thus increasing the ambient neutrino flux and the
thus the magnitude of absorption terms. The asymmetry in the neutron-proton mass
fraction then results in different absorption efficiencies and a net change
in $Y_e$ which always acts in the direction of increasing it (because the
reaction in eqn.~\ref{eq:ghem} occurs more frequently). Figure~\ref{f:delta-ye_mdisk}
shows that at high disk masses, absorption dominates the evolution of $Y_e$ in the
high-compactness models shown. Table~\ref{t:trajectory_integrals} shows
that this trend is also present in models with the lowest 
compactness (b03d01-b03d30), with an even stronger effect of absorption on $Y_e$ 
than in the high compactness sequence.

\begin{figure}
\includegraphics*[width=\columnwidth]{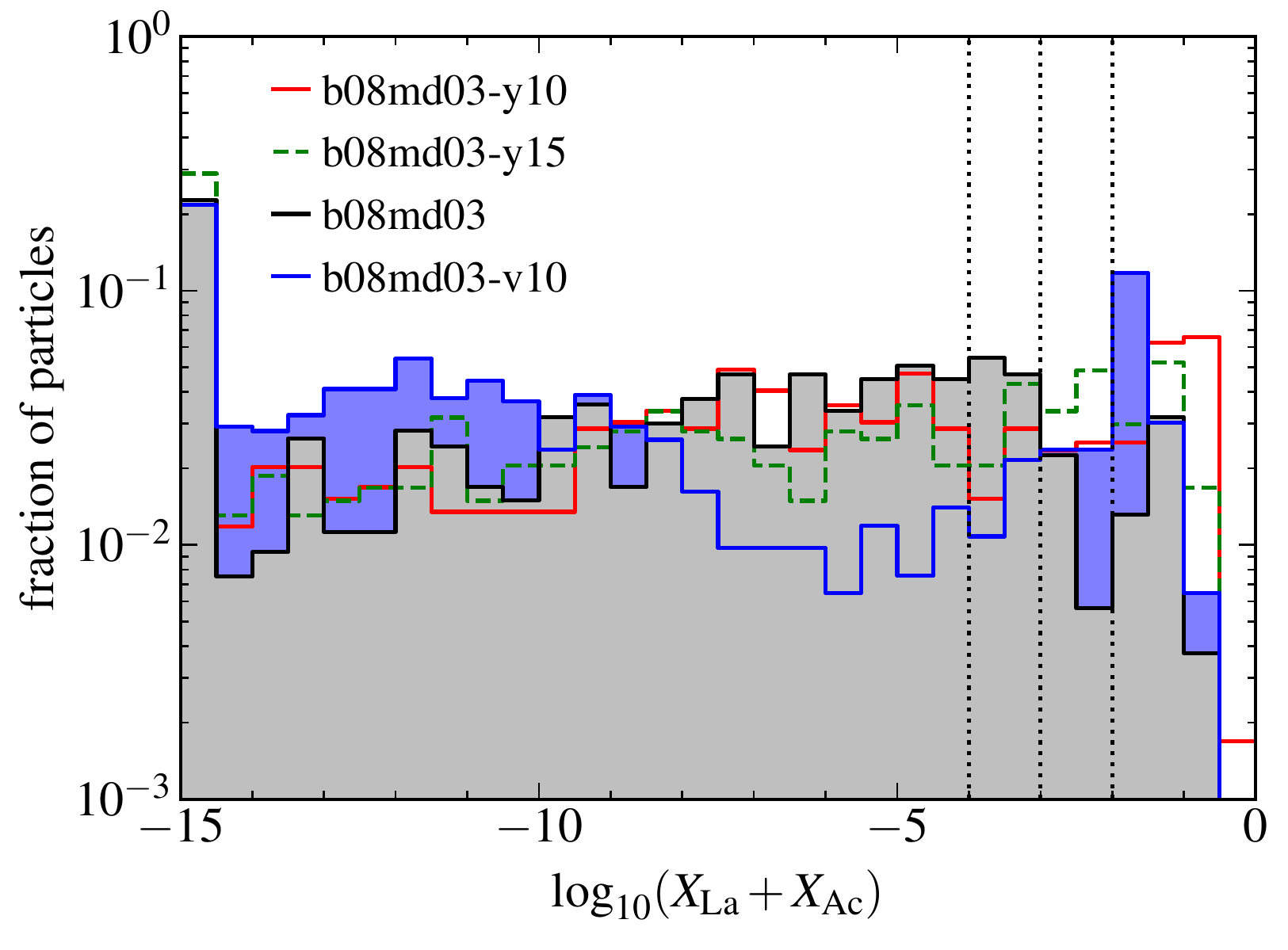}
\caption{Distribution of lanthanide and actinide mass fractions at 1 day, for
nuclear-network-processed particles from models b08d03 (baseline), b08d03-y10 and b08d03-y15 
(varying initial $Y_e$), and b08d03-v10 (high viscosity), as labeled. The bin size is the same is
in the top panel of Figure~\ref{f:hist_lan+ac_abundances}. The lowest bin contains all the particles
with $X_{\rm La}+X_{\rm ac} < 10^{-14.5}$.}
\label{f:hist_lan+ac_ye-visc}
\end{figure}

Models with lower initial $Y_e = \{0.10,015\}$ eject a higher proportion 
of lanthanides and actinides,
as expected, although the change is at a $\lesssim 10\%$ level by mass relative to 
our baseline parameters (Figure~\ref{f:hist_lan+ac_ye-visc}). 
While the $Y_e$ distributions extend to slightly lower electron fractions than
the default models, the bulk of the outflow has $Y_e > 0.2$ in all cases.
Table~\ref{t:trajectory_integrals} shows that these changes are driven
primarily by neutrino emission processes, which adjust
to provide the required change in $Y_e$ toward equilibrium. 

Models with higher viscosity have a similar average $Y_e$ than their low-viscosity counterparts,
but a significantly higher fraction of material rich in lanthanide and actinides. 
The electron fraction distribution of model b08d03-v10 has a tail to low $Y_e$ that
extends well below the lower limit of the distribution of model b08d01. Figure~\ref{f:hist_ye_entropy_b08} 
shows that the material with the lowest $Y_e$ is ejected at the earliest times
in the high-viscosity model, in line with the general expectation that the faster the
evolution of a disk, the stronger the sensitivity of the outflow composition to initial conditions.
This is consistent with the results of GRMHD simulations, which show even stronger memory of
initial conditions given their faster evolution; \citealt{F19_grmhd}.

\begin{figure}
\includegraphics*[width=\columnwidth]{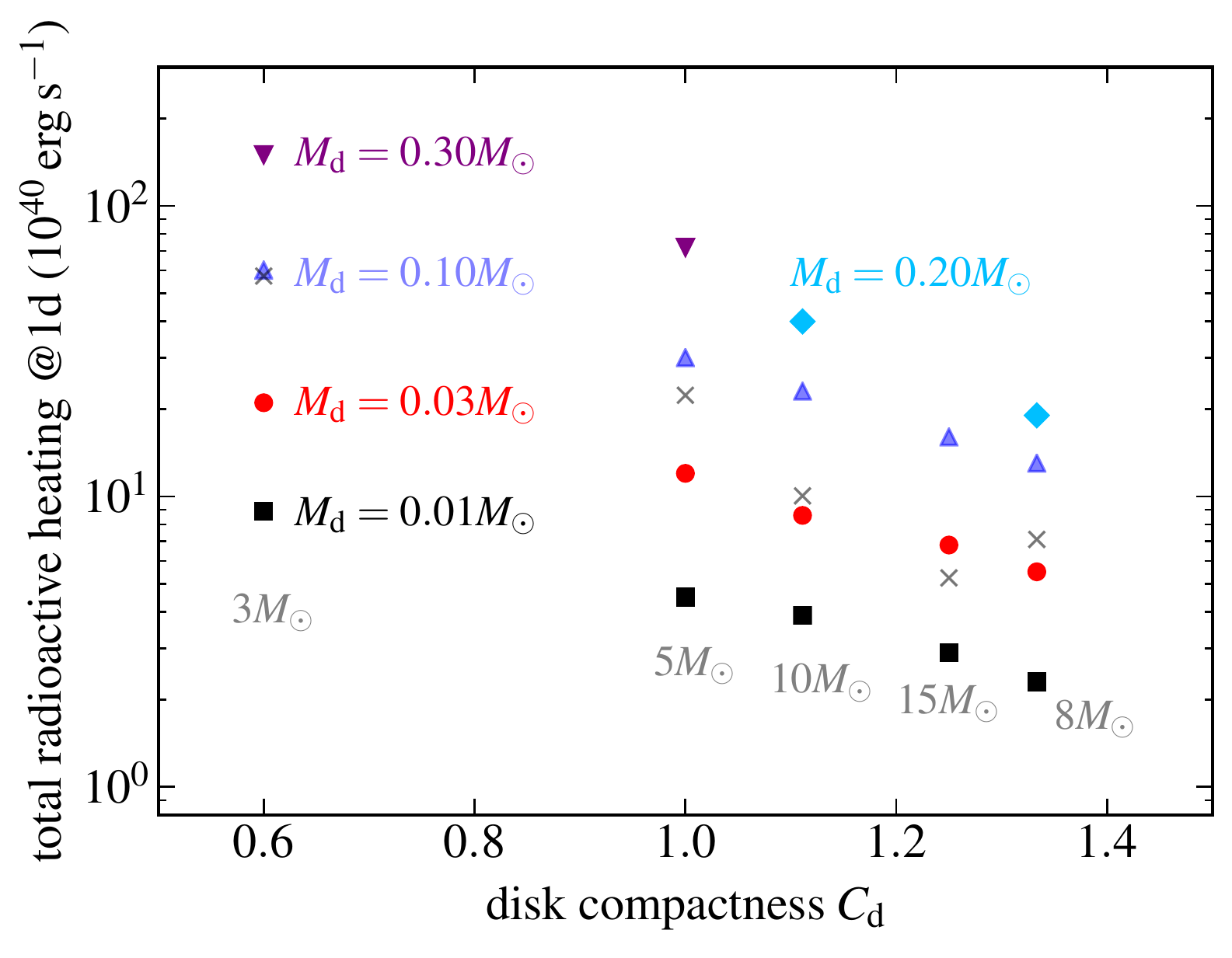}
\caption{Total radioactive heating luminosity at 1 day as a function of disk compactness 
(eq.~\ref{eq:disk_compactness_def}), for various disk masses, as labeled. 
The BH mass is shown in gray under each symbol column. The crosses denote the heating
rates interpolated to the median disk masses from Figure~\ref{f:mdisk_initial}.
The total radioactive heating rate is an upper limit to the bolometric luminosity of the
kilonova, being subject to thermalization efficiency and radiative transfer effects.}
\label{f:lum_compactness}
\end{figure}

%-----------------------------------------------------------------------------------
\subsection{Implications for EM counterpart searches}
\label{s:em_counterparts}

The key question we are interested in is how does the disk outflow
contribute to a kilonova transient. The answer depends on the amount of
mass ejected and its velocity, its composition, as well as how efficiently
does the radioactive heating thermalize (e.g., \citealt{metzger_2019}). 

The mass ejected, and the composition to a lesser extent, determines how much radioactive 
power is available for a kilonova. Figure~\ref{f:lum_compactness} shows the total
radioactive heating luminosity at 1 day as a function of compactness parameter $C_{\rm d}$
and initial disk mass. For fixed compactness, the total radioactive heating is proportional
to the ejected mass, since the average radioactive heating per unit mass is close 
to $2\times 10^{10}$\,erg\,g$^{-1}$\,s$^{-1}$ for most models, given their similar composition.
The dependence of ejected fraction on compactness results in an additional variation of
a factor $\sim 5$ between the least and most compact models, for fixed disk mass.

The raw radioactive heating at 1 day ranges from $2\times 10^{40}$\,erg\,s$^{-1}$ for the
lightest and most compact disk (b08d01) to $1.5\times 10^{42}$\,erg\,s$^{-1}$ for the heaviest
disk in the least compact configuration (b03d30). Thermalization efficiency can result in a
decrease in these values by a factor $\sim 2$ \citep{barnes_2016,waxman_2019,kasen_2019,hotokezaka_2020}, 
while radiative transfer effects (dependent
on the opacity and thus on composition) set whether this power can readily escape the ejecta
or is trapped until later times. Table~\ref{t:results} shows that the radioactive power
at 1 week is about $10$ times smaller than that at 1 day for most models.

If we take the median disk masses from Figure~\ref{f:mdisk_initial} as representative values
for each BH mass and interpolate the ejected masses from Table~\ref{t:results}, we obtain disk 
outflow masses $\{2.00, 0.70, 0.22, 0.25, 0.14\}\times 10^{-2}M_\odot$ for BH masses 
$\{3, 5, 8, 10, 15\}M_\odot$, respectively. 
These values are subject to
enhancement by a factor $\lesssim 2$ 
if GRMHD effects were to be included \citep{F19_grmhd,christie2019}. 

In the case of GW190425, for which our lowest BH mass model would be applicable, the median disk
outflow mass would be similar to the total ejecta from GW170817 within a factor of two, and hence 
it would have been detected with good sky coverage \citep{kyutoku2020}. A BH-NS merger with a 
low-mass BH and high-mass NS is most efficient at tidally disrupting the NS
and most inefficient at producing dynamical ejecta, with most mass ejection coming
from the disk \citep{foucart_2019}. In contrast, two massive NS 
that collapse promptly to a BH are the least efficient configuration for ejecting mass
and forming a disk (e.g., \citealt{radice_2018}, but see
\citealt{kiuchi_2019,bernuzzi_2020} for the case of an asymmetric 
mass ratio NS-NS merger generating a more massive BH accretion disk than a symmetric binary).

Regarding the BH-NS merger candidate GW190814, which had a much smaller localization area
and deeper EM coverage relative to other events, constraints on the
total mass ejection are in the range $0.02-0.1M_\odot$ depending on viewing angle, composition, and distance.
(e.g., \citealt{andreoni_2020,kawaguchi_2020b,morgan_2020,vieira_2020}). Our results indicate that, with the
exception of a very low-mass BH and/or very high disk masses, most BH-NS merger systems would not have 
generated sufficient disk outflow for a detectable kilonova.

An additional factor influencing the kilonova appearance is the relative masses of the disk
and the dynamical ejecta. In a BH-NS merger, the dynamical ejecta is produced along the
equatorial plane, and it blocks only part of the viewing angles. It is expected to be very
rich in lanthanides and thus have a high opacity that blocks the light from the disk
toward most equatorial directions. The results of \citet{FQSKR-15} and \citet{fernandez_2017} -- who studied the 
combined long-term evolution of these two components mapped from dynamical merger simulations --  
show that the bulk of fallback accretion mixes in with the disk before weak freezout occurs and the 
disk outflow is launched. The net outflow has a very similar composition as the dynamical ejecta  and
disk when evolved separately, but with an added component that has intermediate electron fraction values.
The expected kilonova color has an important dependence on viewing angle, as the bluer disk emission
would only be detectable 
from directions not obscured by the dynamical ejecta 
(see, e.g., \citealt{bulla_2019,darbha_2020,kawaguchi_2020a,korobkin_2020} for more recent work on 
viewing angle dependencies of different ejecta configurations).

In hydrodynamic simulations of BH accretion disks, the spatial dependence of the composition 
is quite generic, with the highest $Y_e$ material being ejected first along intermediate latitudes, 
and then wrapping around the outermost edge of the disk outflow \citep{FKMQ14}. This configuration
would guarantee the existence of a blue spike at early times (albeit a faint one) if the merger remnant
is viewable from polar latitudes. All of our models are such that at least $50\%$ of the disk outflow
is lanthanide-poor, and in some cases this fraction can reach even $100\%$ (Table~\ref{t:results}).
GRMHD models show, however, that magnetic fields begin to eject matter much earlier than 
weak freezout, thus adding material that has not been sufficiently processed by 
weak interactions from the initial post-merger composition 
(e.g., \citealt{siegel_2017a,F19_grmhd,miller2019,christie2019}). This will likely
increase the lanthanide-rich fraction at the leading edge of the disk outflow and thus
modify the color of the disk-contributed kilonova.

%%%%%%%%%%%%%%%%%%%%%%%%%%%%%%%%%%%%%%%%%%%%%%%%%%%%%%%%%%%%%%%%%%%%%%%%%%%%
\section{Summary and Discussion}
\label{s:summary}

We have performed axisymmetric hydrodynamic simulations of the viscous 
evolution of accretion disks formed in BH-NS mergers. Our models include
the effects of neutrino emission and absorption on the outflow composition,
and the results are post-processed with a nuclear reaction network for a
more precise diagnostic of the nucleosynthesis yield. These hydrodynamic
models provide a lower-limit to the amount of mass in the disk outflow 
relative to three-dimensional GRMHD models, and hence provide a lower-limit to the
contribution of disk outflows to the kilonova transient. Our simulations cover,
for the first time, a large fraction of the plausible parameter space of BH
and disk masses expected from these mergers (Figure~\ref{f:mdisk_initial}). 
Our main results are the following:
\newline

\noindent
1. -- The fraction of the initial disk mass ejected as an unbound outflow
      has an approximately inverse linear scaling with the initial 
      compactness of the disk, and can vary by a factor of $\sim 4$ 
      (Figure~\ref{f:mass_ejection} and Table~\ref{t:results}).
      While this dependence on compactness was implicit in previous work,
      this is the first time that it is systematically probed over a wide
      parameter space.
      \newline

      \noindent 
      The origin of this dependence can be traced back to the earlier onset
      of accretion in more compact disks, as they are located closer to the
      ISCO. Compared to lower compactness disks, a higher fraction of the
      initial mass is accreted by the time weak interactions freeze out and
      the outflow is launched (Figure~\ref{f:mej_lum_time}). 
      \newline

\noindent 
2. -- At constant compactness, the fraction of the disk mass ejected 
      decreases for higher disk masses. This effect is
      related to the longer time to reach weak freezout in more
      massive disks, which delays the onset of mass ejection to a time when
      more mass has been accreted to the BH (Figure~\ref{f:mej_lum_time}).
      The dependence on initial disk mass is weaker for higher compactness
      systems (Figure~\ref{f:mass_ejection} and Table~\ref{t:results}).
      The initial density and entropy of the disk are the key variables regulating this
      mass dependency of the ejection efficiency (\S\ref{s:mass_ejection}).
      \newline

\noindent 
3. -- The disk outflow is more lanthanide/actinide-poor 
      for higher disk masses (Figure~\ref{f:hist_lan+ac_abundances}), at constant compactness
      (this trend has also been found in previous studies).
      This effect can be traced back to neutrino absorption becoming more important relative
      to emission in increasing $Y_e$ for more massive disks 
      (Figure~\ref{f:delta-ye_mdisk} and Table~\ref{t:trajectory_integrals}).
      Stronger absorption counteracts the action of neutrino emission in lowering $Y_e$ given the lower entropy
      of the outflow from more massive disks (Figure~\ref{f:hist_ye_entropy_b08}).
      \newline

      \noindent
      While our disk outflows are $50-100\%$ lanthanide-poor by mass, magnetically
      driven contributions -- not included here -- can add a significant amount of 
      lanthanide-rich matter, hence the net composition of disk outflows requires 
      simulations with more complete physics for reliable predictions.
      \newline

\noindent
4. -- The ejected fraction and final composition are sensitive to the viscosity parameter of the
      disk, as known from previous work, with more mass ejected, at higher velocities, and with a higher 
      lanthanide-rich fraction for higher viscosity parameter (Table~\ref{t:results}). The 
      most neutron-rich material is produced at the earliest times in the simulation (Figure~\ref{f:hist_ye_entropy_b08})
      and is thus related to the shorter evolution time of these disks.
      \newline

\noindent
5. -- In most cases, the initial $Y_e$ of the disk has a negligible effect on the amount of mass ejected,
      with the exception of massive disks in low-compactness systems, where the effect modifies the
      ejection efficiency by a few percent of the disk mass (similar to numerical resolution). Hence,
      the ejected fraction is mostly
      robust to variations in the initial composition (Table~\ref{t:results}).
      The final composition does depend on the initial conditions, with corrections of the order
      of $10\%$ to the fraction of the outflow mass that is lanthanide-rich (Figure~\ref{f:hist_lan+ac_ye-visc})
      \newline

\noindent
6. -- The range of ejecta masses from the disk outflow can result
      in a range of two orders of magnitude in raw radioactive luminosity over the BH-NS parameter 
      space probed (Figure~\ref{f:lum_compactness}). Except for very low-mass BHs and/or very
      high disk masses, most BH-NS mergers should generate disk outflows that are below constraints 
      for the total ejecta mass from the BH-NS merger candidate GW190814 (\S\ref{s:em_counterparts}).
      \newline

Our results are consistent with previous hydrodynamic models of BH accretion disks. Model t-a80-hr from
\citet{FKMQ14} is equivalent to our model b03d01 but evolved until $3,000$ orbits. By that time,
the total and unbound (positive energy) mass ejected in their model are $19\%$ and $12\%$, respectively,
while here we obtain $21\%$ and $14\%$, respectively. The $\sim 10\%$ difference can be attributed to
the lower ambient and floor of density used here, and on improvements in the neutrino implementation
as reported in \citet{lippuner_2017}. Similarly, model Fdisk of \citet{fernandez_2017} falls
in between models b08d03 and b08d10 in terms of compactness, but has a slightly higher BH spin (0.86).
The higher ejected fraction in their model ($8\%$) can be partially accounted for by the higher value
of the BH spin, and also by a more extended initial density distribution with radius 
in the torus mapped from the merger simulation (covering a wider range in compactness)

We have made specific choices for the disk entropy (\S\ref{s:initial_conditions}), which can have implications for
the sensitivity of the ejected fraction to initial disk mass. Similarly, the choice of
viscosity parameter is on the low end of values that bracket the amount of angular momentum transport seen in 
GRMHD simulations of equivalent accretion disks \citep{F19_grmhd}. More realistic
values for these two parameters must come from direct mapping of the outcome of
GRMHD simulations of BH-NS mergers that include neutrino transport, where the magnetic field
geometry and strength replaces the viscosity parameter. 
A direct mapping would also avoid the need to make well-motivated but ultimately
arbitrary choices for an initial torus radius, which is required by an equilibrium initial condition
and directly enters the compactness parameter (equation~\ref{eq:disk_compactness_def}).
Mapping from merger simulations would also inform the
initial $Y_e$ distribution which, while not crucial for determining the amount of mass ejected,
has implications for the outflow composition at the level of detail needed for accurate
predictions of kilonovae and nucleosynthesis yields. Compact object merger simulations
that account for both MHD and neutrino effects are few and only implement the latter via leakage
schemes (e.g., \citealt{nielsen_2014,most_2019}). Simulations with more advanced (e.g., M1) neutrino
transport do not yet include MHD effects. 

Our results suggest that GRMHD simulations carried out with no neutrino absorption (e.g., 
\citealt{siegel_2018,F19_grmhd,christie2019}) can give reasonable $Y_e$ distributions
only for very low mass disks $< 0.01M_\odot$, for which the absorption contribution to
$Y_e$ is subdominant.
For the configuration 
studied by \citet{siegel_2018}, \citet{F19_grmhd}, and \citet{christie2019}
($M_{\rm bh}=3M_\odot$ and $M_{\rm d}=0.03M_\odot$),
we find that absorption is already more important than emission in setting the $Y_e$ distribution 
(Table~\ref{t:trajectory_integrals}), in line with the results of \citet{miller2019} who showed 
a significant increase in the $Y_e$ of the outflow when including neutrino absorption. 

Further GRMHD studies of BH accretion disks over a wide region of parameter space are needed to determine
whether the ejected fraction has the same dependence on compactness as in pure hydrodynamic models,
and whether the fraction of the outflow that is lanthanide-rich is significantly
larger than what we find here. Both of these questions are crucial to improve predictions of EM counterparts
to BH-NS and NS-NS sources, and require (1) models evolved for long timescales, (2)
realistic initial field strengths, geometries, entropies, and electron fractions, 
and (3) inclusion of neutrino absorption. Such simulations remain challenging given
current algorithms and computational resources.

\section*{Acknowledgements}

We thank the anonymous referee for helpful comments that improved the presentation
of the paper.
RF acknowledges support from the National Sciences and Engineering Research Council (NSERC)
of Canada through Discovery Grant RGPIN-2017-04286, and from the Faculty of Science at 
the University of Alberta.
FF gratefully acknowledges support from the U.S. National Science Foundation (NSF)
through grant PHY-1806278, from the National Aeronautics and Space Administration (NASA)
through grant 80NSSC18K0565, and from the U.S. Department of Energy CAREER
grant DE-SC0020435.
This work was supported by the U.S. Department of Energy through the Los Alamos
National Laboratory. Los Alamos National Laboratory is operated by Triad National
Security, LLC, for the National Nuclear Security Administration of U.S. Department
of Energy (Contract No. 89233218CNA000001). This work was assigned report number
LA-UR-20-23877.
The software used in this work was in part developed by the U.S. Department of Energy
NNSA-ASC OASCR Flash Center at the University of Chicago.
This research was enabled in part by support
provided by WestGrid (www.westgrid.ca), the Shared Hierarchical
Academic Research Computing Network (SHARCNET, www.sharcnet.ca), Calcul Qu\'ebec (calculquebec.ca),
and Compute Canada (www.computecanada.ca).
Computations were performed on \emph{Graham}, \emph{Cedar}, and \emph{B\'eluga}.
This research also used storage resources of the U.S. National Energy Research Scientific Computing
Center (NERSC), which is supported by the Office of Science of the U.S. Department of Energy
under Contract No. DE-AC02-05CH11231 (repository m2058).
Graphics were developed with {\tt matplotlib} \citep{hunter2007}.

%%%%%%%%%%%%%%%%%%%%%%%%%%%%%%%%%%%%%%%%%%%%%%%%%%
\section*{Data Availability}

The data underlying this article will be shared on reasonable request to the corresponding author.

%%%%%%%%%%%%%%%%%%%% REFERENCES %%%%%%%%%%%%%%%%%%
\bibliographystyle{mnras}
\bibliography{ms} 

%%%%%%%%%%%%%%%%%%%%%%%%%%%%%%%%%%%%%%%%%%%%%%%%%%
\bsp	
\label{lastpage}
\end{document}